\documentclass[conference]{IEEEtran}

\usepackage{graphicx}
\usepackage{lipsum}
\usepackage{hyperref}
\usepackage{cite}
\usepackage{pifont}
\usepackage{enumitem}
\usepackage{subcaption}

\usepackage{cases}
\usepackage{adjustbox}
\usepackage{array, diagbox}
\usepackage[font=small,skip=0pt]{caption}
\usepackage{subcaption}
\usepackage{amsmath}
\IEEEoverridecommandlockouts
\usepackage{atbegshi}% http://ctan.org/pkg/atbegshi
\AtBeginDocument{\AtBeginShipoutNext{\AtBeginShipoutDiscard}}

\begin{document}

%+++++++++++++++++++++++++++++++++++++++++++
\title{\LARGE Spectral Efficiency Maximization for mmWave MIMO-Aided Integrated Sensing and Communication Under Practical Constraints
}
%\pagenumbering{arabic}
\author{\small
Jitendra~Singh, \textit{Graduate Student Member,~IEEE,}
Anand~Mehrotra, \textit{Graduate Student Member,~IEEE,}
Suraj~Srivastava,\\ \textit{Member,~IEEE,}
Aditya~K.~Jagannatham, \textit{Senior Member,~IEEE}
and Lajos~Hanzo, \textit{Life Fellow,~IEEE}
}
\thanks{J. Singh A. Mehrotra and A. K. Jagannatham are with the Department of Electrical Engineering, Indian Institute of Technology Kanpur, Kanpur, UP 208016, India (e-mail: jitend@iitk.ac.in; anandme@iitk.ac.in; adityaj@iitk.ac.in).
}
\thanks{S. Srivastava is with the Department of Electrical Engineering, Indian Institute of Technology Jodhpur, Jodhpur, Rajasthan 342030, India (email: surajsri@iitj.ac.in).}
\thanks{L. Hanzo is with the School of Electronics and Computer Science, University of Southampton, Southampton SO17 1BJ, U.K. (e-mail: lh@ecs.soton.ac.uk).}
%\thanks{The work of Aditya K. Jagannatham was supported in part by the Qualcomm Innovation Fellowship; in part by the Qualcomm 6G UR Gift; in part by the Arun Kumar Chair Professorship; and in part by the DST, Govt. of India and –UKRI EPSRC Project Intelligent Spectrum Innovation ICON. The work of S. Srivastava was supported in part by IIT Jodhpur's Research Grant No. I/RIG/SUS/20240043; in part by Anusandhan National Research Foundation's PM-ECRG/2024/478/ENS; and in part by Telecom Technology Development Fund (TTDF) under Grant TTDF/6G/368. S. Srivastava and A. K. Jagannatham jointly acknowledge the funding support provided to ICON-project by DST and UKRI-EPSRC under India-UK Joint opportunity in Telecommunications Research. 
%L. Hanzo would like to acknowledge the financial support of the Engineering and Physical Sciences Research Council (EPSRC) projects under grant EP/Y026721/1, EP/W032635/1, EP/Y037243/1 and EP/X04047X/1 as well as of the European Research Council's Advanced Fellow Grant QuantCom (Grant No. 789028).}

\maketitle
\begin{abstract}
A hybrid transmit precoder (TPC) and receive combiner (RC) pair is conceived for millimeter wave (mmWave) multiple input multiple output (MIMO) integrated sensing and communication (ISAC) systems. The proposed design considers a 
practical mean squared error (MSE) constraint between the desired and the achieved beampatterns constructed for identifying radar targets (RTs). To achieve optimal performance, we formulate an optimization problem relying on sum spectral efficiency (SE) maximization of the communication users (CUs), while satisfying certain radar beampattern similarity (RBPS), total transmit power, and constant modulus constraints, where the latter are attributed to the hybrid mmWave MIMO architecture. Since the aforementioned problem is non-convex and intractable, a sequential approach is proposed wherein the TPCs are designed first, followed by the RCs. To deal with the non-convex MSE and constant modulus constraints in the TPC design problem, we propose a majorization and minimization (MM) based Riemannian conjugate gradient (RCG) method, which restricts the
tolerable MSE of the beampattern to within a predefined limit. Moreover, the least squares and the zero-forcing methods are adopted for maximizing the sum-SE and for mitigating the multiuser interference (MUI), respectively. Furthermore, to design the RC at each CU, we propose a linear MM-based blind combiner (LMBC) scheme that does not rely on the knowledge of the TPC at the CUs and has a low complexity. 
To achieve user fairness, we further extend the proposed sequential approach for maximizing the geometric mean (GM) of the CU's rate. %These compelling properties make it eminently suitable for real-world implementation. 
Simulation results are presented, which show the superior performance of the proposed hybrid TPC and RC in comparison to the state-of-the-art designs in the mmWave MIMO ISAC systems under consideration. 
\end{abstract}

\maketitle
\section{Introduction}
\IEEEPARstart{I}{ntegrated} sensing and communications (ISAC) is garnering enormous interest in both academia and industry due to its hardware and spectrum sharing capability, which improves the spectral efficiency (SE) of the systems \cite{ISAC_mmWave_1_1,ISAC_mmWave_9,jigaung_1}. 
%There are two popular models in ISAC systems, viz., dual functional radar communication (DFRC) systems, wherein the communication and radar systems share the same hardware and communication-radar coexistence. In the later, the radar and communication systems are distinct, while operating in the same frequency band. 
Numerous attractive applications envisaged for ISAC in 6G systems, such as providing backhaul data connectivity for vehicle-to-everything (V2X) communications 
\cite{ISAC_MIMO_23}, assisted beamforming for vehicle to infrastructure (V2I) networks \cite{ISAC_mmWave_10}, connected automated vehicle (CAV) applications\cite{ISAC_mmWave_10}, remote sensing, environmental monitoring, human computer interaction, extended reality (XR) \cite{applications_1} etc.
Moreover, as a further advancement in communication technology and also to exploit some of the hitherto under-used spectral bands, the operating frequencies have been expanded to include the millimeter wave (mmWave) band \cite{ISAC_mmWave_9,ISAC_mmWave_10}. Naturally, mmWave communication technology presents an excellent opportunity for ISAC services in terms of its channel characteristics and signal processing solutions \cite{ISAC_MIMO_23,applications_1}. Hence, this paper investigates mmWave multi-input multi-output (MIMO) ISAC systems, and focuses on the hybrid transmit precoder (TPC) and receive combiner (RC) design to maximize the SE of the communication subsystem. 
\subsection{Prior work and motivation} \label{literature review}
Recently, the waveform design problem of MIMO radar systems has been extensively studied in \cite{ISAC_MIMO_6,ISAC_MIMO_8,ISAC_MIMO_7,ISAC_MIMO_16,ISAC_MIMO_17,ISAC_MIMO_21,ISAC_IRS_1,ISAC_IRS_3}. Specifically, the authors of \cite{ISAC_MIMO_6} have proposed transmit beamforming at the ISAC BS, which aligns the beam towards the radar targets (RTs), while satisfying the signal to interference plus noise ratio (SINR) requirement as the quality of service (QoS) constraint of each communication user (CU) in a multi-user (MU) ISAC MIMO system. 
%The authors of \cite{ISAC_MIMO_6} consider a pair of models based on either separate or shared antennas for the RTs and CUs. They concluded that the latter yields a substantially improved tradeoff between the quality of the radar beampattern and the QoS of CUs in comparison to the former. 
%As a further advance, Wang \textit{et al.} \cite{ISAC_MIMO_8} introduced dedicated signals for sensing, which supported higher degrees of freedom for synthesizing the desired beampattern of the RTs. 
The authors of \cite{ISAC_MIMO_16} have designed an ISAC waveform for a MU MIMO ISAC system for optimizing the weighted performance trade-off between the sensing quality of the RTs and the SINR constraint of the CUs, subject to either total or per-antenna power constraints.
%Specifically, Wei \textit{et al.} \cite{ISAC_MIMO_7} investigated orthogonal frequency division multiplexing (OFDM) MIMO-based ISAC systems and presented a novel metric for sensing and communication, namely mutual information, which is based on an information-theoretic approach. The authors proposed a potent waveform design for maximizing the weighted sum of the mutual information of sensing and communication.
%To investigate the multi-beam transmit beamforming framework, the authors of \cite{ISAC_MIMO_7} employ analog steering antenna arrays for ISAC MU MIMO systems, which are remarkably suitable for integration into the systems relying on the existing time division duplex (TDD) protocol associated with orthogonal frequency division multiplexing (OFDM).  
%Furthermore, to improve the performance of ISAC MIMO systems, Wen \textit{et al.}, \cite{ISAC_MIMO_17}, proposed a hybrid linear-nonlinear TPC scheme acting as the transmit beamformer, which was shown to yield improved performance over the conventional linear TPC in terms of the accuracy of radar target (RT) detection. 
Aldayel \textit{et al.} \cite{ISAC_MIMO_21} consider the problem of waveform design in MIMO radar systems, which maximizes the SINR of the communication users (CUs) under the practical constant modulus and radar beampattern similarity (RBPS) constraints. 

However, the authors of \cite{ISAC_MIMO_6,ISAC_MIMO_8,ISAC_MIMO_7,ISAC_MIMO_16,ISAC_MIMO_17} conceived transmit beampattern design algorithms for ISAC MIMO systems, where a separate RF chain is required for each antenna. But, such an architecture is inefficient for mmWave MIMO systems relying on large scale antenna arrays due to the requirement of a large number of RF chains, which increases both the cost and energy consumption. For reducing the number of RF chains, the recently proposed hybrid architecture \cite{HBF_8,HBF_6,HBF_7,HBF_9} has been shown to be immensely well-suited for mmWave MIMO systems, where the signal processing procedures are partitioned between the digital baseband and analog RF domains. 
%The authors of \cite{HBF_3} proposed innovative hybrid TPC design techniques for mmWave MIMO systems, where the BB and RF TPCs are designed using an alternating minimization approach in which the Riemannian manifold optimization principle is adopted to design the RF TPC, taking into account the CM constraint of each element. As a further advance, the authors of \cite{HBF_1} designed a hybrid TPC and RC framework for MU mmWave MIMO systems, wherein various innovative interference cancellation methods are proposed for maximizing the sum-SE of the system. To ensure fairness, Yu \textit{et al.} \cite{HBF_6} proposed a hybrid TPC design, which maximizes the geometric mean of user rates in a MU mmWave system. 
To explore the pros and cons of ISAC systems in the mmWave frequency band, various recent studies \cite{ISAC_mmWave_1,ISAC_mmWave_3,ISAC_mmWave_1_1,ISAC_mmWave_12_1,ISAC_mmWave_15,ISAC_mmWave_4,ISAC_mmWave_5,ISAC_mmWave_6,ISAC_mmWave_14,ISAC_mmWave_13,ISAC_mmWave_12,ISAC_mmWave_IRS_1,ISAC_mmWave_IRS_2,ISAC_mmWave_IRS_3,new_13,new_12,ISAC_mmWave_35} have proposed hybrid TPC/RC transceivers for mmWave MIMO ISAC systems. 
In their pioneering work Liu \textit{et al.,} \cite{ISAC_mmWave_1}, have conceived hybrid TPC designs for both fully connected and sub-arrayed connected MIMO radar architectures operating in mmWave MIMO ISAC systems. Their workoptimized the RF and baseband TPCs using a triple alternating minimization (TAltMin) method to strike a trade-off between the performance of the CUs and the RTs.
%In their pioneering work Liu \textit{et al.,} \cite{ISAC_mmWave_1}, have conceived hybrid TPC designs for both fully connected and sub-arrayed MIMO radar architectures operating in mmWave MIMO ISAC systems. They optimized the RF and baseband TPCs using a triple alternating minimization (TAltMin) method to accomplish a trade-off between the performance of the CUs and the RTs. }
Qi \textit{et al.} \cite{ISAC_mmWave_3} minimized the error between the radar beampattern and the desired transmit beamformer, incorporating minimum SINR constraints for each CU and total power constraints to design the hybrid TPCs for mmWave MIMO ISAC systems.
%Qi \textit{et al.} \cite{ISAC_mmWave_3} minimized the error between the radar beampattern and the desired transmit beamformer, incorporating minimum SINR constraints for each CU and total power constraints to design hybrid TPCs for mmWave MIMO ISAC systems.}
Furthermore, Wang \textit{et al.} \cite{ISAC_mmWave_12_1} minimize the Cramér-Rao bound (CRB) of the direction of arrival (DOA) estimation of the RTs, while considering the minimum SINR requirement of the CUs in a mmWave ISAC system.
%Dong \textit{et al.} \cite{ISAC_mmWave_2} considered a partially connected hybrid architecture at the BS and proposed a triple alternating minimization technique for designing a hybrid beamforming scheme to accomplish the joint sensing and communication tasks. 
Moreover, to investigate the implementation and benefits of full-duplex (FD) communication in mmWave MIMO ISAC systems, Islam \textit{et al.} \cite{ISAC_mmWave_4} consider a BS having a hybrid architecture capable of communicating with the CUs in the downlink and also concurrently performing sensing of the RTs by exploiting the same reflected signals. 
As a further advance, the authors of \cite{new_13} adopted a correlated communication-sensing channel model to analyze the performance of a multibeam mmWave-enabled ISAC system. 
\begin{table*}[t!]
    \centering
    \caption{Contrasting our novel contributions to the literature of mmWave MIMO ISAC systems} \label{tab:lit_rev}
    \begin{adjustbox}{width=\linewidth}
\begin{tabular}{|l|c|c|c|c|c|c|c|c|c|c|}
    \hline
&\cite{ISAC_mmWave_1_1}   &\cite{ISAC_MIMO_6,ISAC_MIMO_8,ISAC_MIMO_16}  &\cite{ISAC_MIMO_17,ISAC_MIMO_21}     &\cite{ISAC_mmWave_1,ISAC_mmWave_12_1,ISAC_mmWave_6,ISAC_mmWave_14} &\cite{ISAC_mmWave_3}  &\cite{ISAC_mmWave_15} &\cite{ISAC_mmWave_13} &\cite{ISAC_mmWave_12} &\cite{GM_1,GM_2} & Proposed \\ [0.5ex]
 \hline
ISAC system &\checkmark    &\checkmark   &\checkmark    &\checkmark  &\checkmark &\checkmark &\checkmark  &\checkmark  &  &\checkmark\\
\hline
mmWave MIMO &\checkmark    &    &     &\checkmark &\checkmark &\checkmark &  &\checkmark &  &\checkmark\\
\hline
%Fully-connected structure &\checkmark    &  &  &  &  &   &\checkmark  &\checkmark &   &  & &\checkmark &\checkmark &\checkmark  &\checkmark & &\checkmark\\
%\hline
Hybrid beamforming &\checkmark   &    &    &\checkmark &\checkmark  &\checkmark &\checkmark  &\checkmark &  &\checkmark\\
\hline
Multiple CUs &\checkmark  &\checkmark  &\checkmark    &\checkmark &\checkmark  &\checkmark  &\checkmark &\checkmark  &\checkmark  &\checkmark\\
 \hline
Multiple RTs &\checkmark  &\checkmark   &\checkmark    &\checkmark &\checkmark  & &\checkmark  &\checkmark  &  &\checkmark\\
 \hline
Sum-SE maximization &\checkmark  &    &     & &\checkmark & &\checkmark  &\checkmark  &  &\checkmark\\
 \hline
%RBPS
%&  &     &\checkmark   &  &\checkmark     &  &  &\checkmark \\
% \hline
%Sequential method
%&  &    &    & &  & & &\checkmark \\
% \hline
Two-stage hybrid TPC
&  &    &   &  &\checkmark &\checkmark   & & &  &\checkmark \\
 \hline
%MM 
%&   &  &  &  & &  &  &\checkmark &  &  & &  & &  &  &\checkmark &\checkmark \\
% \hline
%RCG &  &\checkmark &  &\checkmark  & &   &  &\checkmark &  & &\checkmark  &  &\checkmark &  &\checkmark &  &\checkmark\\
% \hline
%ZF &\checkmark  &\checkmark &\checkmark &  & &  &   &\checkmark &  &  & &  & &\checkmark   &  & 
% &\checkmark\\
% \hline
{\bf GM-SE maximization}
&  &   & & &  & & &  &\checkmark  &\checkmark \\
 \hline
{\bf MSE as RBPS}
&  &   & & &  & & & &  &\checkmark \\
 \hline
{\bf MM-based RCG} &  &   &  & & &  &  & & &\checkmark\\
 \hline
{\bf MM-based blind RC}
&  &    & &  &  & & &  &  &\checkmark\\
 \hline
\end{tabular}
\end{adjustbox}
\end{table*}
%Liu \textit{et al.} \cite{ISAC_mmWave_14} consider the application of mmWave MIMO ISAC systems in a V2X scenario. The authors therein proposed a single-target multi-beam (STMB) scheme for enhancing both the accuracy of velocity and range estimation by allocating multiple radar beams to a certain RT.   
Furthermore, to maximize the sum-rate of the CUs, the authors of \cite{ISAC_mmWave_13,ISAC_mmWave_15,ISAC_mmWave_12,new_12} investigated the hybrid TPC design of mmWave MIMO ISAC systems considering the radar beam patter similarity (RBPS) constraint. Specifically, the authors of \cite{ISAC_mmWave_13,ISAC_mmWave_15} adopted the alternating direction method of multipliers (ADMM) and the weighted mean-square error minimization (WMMSE) principles for single-CU and multi-CU scenarios, respectively. Yu \textit{et al.} \cite{ISAC_mmWave_12} proposed a fast Riemannian manifold optimization (FRMO) and adaptive particle swarm optimization (APSO) algorithms for single and multiple CU scenarios, respectively. Gong \textit {et al.} \cite{new_12} proposed a hybrid TPC design aimed at maximizing the weighted sum rate of the CUs, while constraining the CRB for radar target angle estimation. The authors therein employed the RCG and successive convex approximation (SCA) methods to effectively solve the optimization problem formulated.
%Gong \textit {et al.} \cite{new_12} proposed a hybrid TPC design aimed at maximizing the weighted sum rate of the CUs while constraining the CRB for radar target angle estimation. The authors therein employed Riemannian conjugate gradient (RCG) and successive convex approximation (SCA) methods to effectively solve the formulated optimization problem. }

There is a paucity of studies that have investigated sum-rate maximization for communication users (CUs) while incorporating realistic radar beampattern similarity (RBPS) constraints in ISAC systems; only a few works such as \cite{ISAC_mmWave_13,ISAC_mmWave_15,ISAC_mmWave_12} have explored this topic. These approaches formulate the RBPS constraint by comparing the designed hybrid transmit beamformer with an optimal radar-only beamformer, which is typically derived using full channel state information (CSI) of radar targets—including Doppler shifts, delays, and path gains. However, acquiring such full CSI is extremely challenging in mmWave ISAC systems due to large antenna arrays and highly directional, sparse propagation environments.
By contrast, our work introduces a more practical RBPS formulation that leverages the mean squared error (MSE) between the desired and designed transmit beampatterns. In this context, the desired beampattern can be constructed using only the angular information (e.g., target direction estimates), without requiring full CSI. This makes our proposed metric particularly well-suited for mmWave ISAC implementations, especially in real-world scenarios such as vehicular networks, smart surveillance, and human activity recognition, where estimating full radar CSI, including the Doppler, the path delays \& gains, is infeasible.

Moreover, only the authors of \cite{ISAC_mmWave_15} consider RC design for sum-SE maximization in mmWave MIMO ISAC systems. The authors therein jointly designed the TPC and RC at the ISAC BS, which requires feedback of the intermediate results, which imposes severe overheads on ISAC implementations. 
To overcome these challenges and address the knowledge gaps in existing research, in this work, we conceive novel hybrid TPC and RC design techniques for mmWave MIMO ISAC systems. The proposed schemes maximize the sum-SE, while adhering to the RBPS, to the total transmit power and to the mmWave MIMO hybrid TPC constraints. 
Furthermore, unlike conventional ISAC approaches that require explicit feedback of the TPCs from the CUs to optimize sensing and communication performance, our proposed RC design operates without such feedback. This eliminates the need for TPC exchange between the ISAC BS and the CUs, significantly reducing signaling overhead and enhancing the practicality of ISAC systems for real-world deployments.
%Thus, the approach advocated in this paper is substantially different in comparison to the related works \cite{ISAC_mmWave_13,ISAC_mmWave_15,ISAC_mmWave_12}. 
Furthermore, the existing literature of mmWave ISAC systems has not as yet explored the principle of rate fairness for the CUs, which is also a key performance metric. To elaborate briefly, maximizing the sum rate typically assigns the most resources to the CUs having the best channel, while assigning a near-zero rate to the CUs having low channel quality, especially as the radar performance improves. However, in order to ensure rate fairness, significant research efforts have been dedicated to the beamforming design, stopping that of the more challenging joint communications and sensing optimization, focusing either on maximizing the minimum CU rate or the GM rate of the CU \cite{GM_1,GM_2,new_103}. 
%The study \cite{Without_ISAC_1} addresses the maximization problems of SR, MR, and GMR in massive MIMO systems. 
The authors of \cite{GM_1} proposed a cutting-edge HBF design by solving a max-min rate (MMR) optimization problem in a mmWave system. Moreover, Tuan \textit{et al.} \cite{new_103} considered an RIS-aidedd wireless system and jointly optimized the active and passive beamformers by maximizing the GM rate of the CUs. Furthermore, the novel transformations of the objectives in \cite{GM_1,GM_2} were conceived using the theory of majorization-minimization (MM), and subsequently, closed-form expressions were derived for the optimal solutions, which renders these studies potent in practical deployments.
Compared to MMR optimization, which prioritizes the worst-case user, maximization of GM rates balances rate fairness and overall performance. In mmWave ISAC systems with directional beams and blockages, resource domination by a single user is prevented at the excessive cost of improving weaker user rates. This enhances the system's overall spectral efficiency and mitigates performance degradation seen in strict MMR formulations, making it a practical choice for ISAC deployments. Motivated by these facts, we further maximize the geometric mean (GM)-SE of the system \cite{GM_1,GM_2,new_103}, which leads to a similar rate for all CUs.
Our novel contributions are boldly and explicitly contrasted to the existing literature in Table  \ref{tab:lit_rev} and are further described next.

%Authors of \cite{ISAC_mmWave_IRS_1} consider the intelligent reflecting surface (IRS)-aided mmWave MIMO ISAC systems and jointly optimize the radar signal covariance matrix, communication beamforming vectors and IRS reflection matrix, which maximizes the sum rate of the communication users under the matching of the radar beampattern and the total power transmit power constraints. Authors of \cite{ISAC_mmWave_IRS_2} investigated the IRS-assisted mmWave single input and multiple output (SIMO) systems, wherein the location sensing and data transmission occurs in the same time-frequency resource block. A transmission protocol of the system is developed in which the location is sensed at the IRS by using the communication signal, which is further used by the BS to design the beamforming for communication purpose. Furthermore, authors in \cite{ISAC_mmWave_IRS_3} take the destructive interference due to the echo of the communication users on the sensing performance in IRS-aided mmWave MIMO ISAC systems and proposed a novel feedback based beam training framework to ensure both the sensing and communication performance.

\subsection{Contributions of this work}\label{contributions}
\begin{enumerate}
    \item We commence by developing a system model for the mmWave MIMO ISAC downlink, which serves multiple CUs while sensing multiple RTs. Furthermore, we consider an MSE-based RBPS constraint in the sum-SE maximization problem of an mmWave MIMO ISAC system.
    Additionally, the total transmit power and hardware constraints imposed by the mmWave hybrid MIMO architecture are also taken into consideration.
    \item In order to solve the above-mentioned challenging problem, we first decouple the hybrid transceiver design optimization problem into that of the capacity approaching and sensing-optimal hybrid TPC as well as the MMSE-optimal hybrid RC design sub-problems. Subsequently, to design the hybrid TPC at the ISAC BS, we propose a two-stage hybrid TPC design, in which the RF TPC and stage-1 baseband TPC are designed jointly for maximizing the sum-SE of the system while meeting the RBPS requirements of the RTs. Next, in the second stage, the baseband TPC is designed for mitigating the MUI emanating from the CUs and the RTs. 
    \item To jointly design the RF and baseband TPCs in the first-stage, we propose a MM-based RCG (RMCG) algorithm, which develops a family of convex surrogate functions for the non-convex RBPS function and thereafter obtains the RF TPC by invoking the RCG algorithm. Furthermore, the stage-1 and stage-2 baseband TPCs are determined using the least square and ZF methods, respectively.
    \item To design the RC at each CU, we formulate the RC optimization problem based on the MMSE principle, which is then solved using an innovative linear
    MM-based blind combiner (LMBC) algorithm that does not require feedback of the TPCs of the CUs.  Furthermore, to handle the user unfairness in the sum-SE maximization, we extend the proposed TPC/RC design to maximize the GM-SE of the system.
    \item The performance of the proposed hybrid TPC and RC designs is characterized via simulations and also compared to the pertinent benchmarks, which demonstrates the efficiency of the proposed methods.
\end{enumerate}
\subsection{Notation}\label{notation}
Boldfaced uppercase letters and boldfaced lowercase letters represent matrices, and vectors, respectively;
The $i$th column, the $(i,j)$th element, and the Hermitian of matrix $\mathbf{A}$ are denoted by $\mathbf{A}{(:,i)}$, $\mathbf{A}{(i,j)}$, and $\mathbf{A}^H$, respectively; the Hermitian and conjugate transpose of a matrix $\mathbf{A}$ are denoted by $\mathbf{A}^H$ and $\mathbf{A}^*$, respectively; $\left\vert\left\vert \mathbf{A} \right\vert\right\vert_F$ denotes the Frobenius norm of $\mathbf{A}$, whereas $\left\vert\mathbf{A}\right\vert$ represents its determinant; $\mathbf{a}(i)$ denotes the $i$th element of vector $\mathbf{a}$, and
$\left\vert\mathbf{a}\right\vert$ and $\angle\mathbf{a}$ represent the magnitude and the phase vectors of $\mathbf{a}$;
${\cal D}(\mathbf{a})$ denotes a diagonal matrix with vector $\mathbf{a}$ on its main diagonal; $\text{vec}(\mathbf{A})$ is the vectorization of the matrix $\mathbf{A}$ by stacking its columns; $\mathbf{A} \odot\mathbf{B}$ is the Hadamard product of $\mathbf{A}$ and $\mathbf{B}$; $\nabla f\big(\mathbf{a}\left(i\right)\big)$ denotes the gradient vector of function $f\big(\mathbf{a}\left(i\right)\big)$ at the point $\mathbf{a}\left(i\right)$;
The expectation operator is represented as $\mathbb{E}[\cdot]$; the real part of a quantity is denoted by $\Re\{\cdot\}$; ${\mathbf I}_M$ denotes an $M \times M$ identity matrix; the symmetric complex Gaussian distribution of mean $\mathbf{a}$ and covariance matrix $\mathbf{A}$  is represented as ${\cal CN}(\mathbf{a}, \mathbf{A})$.
\section{System Model}\label{System Model}
\begin{figure}
    \centering
    \includegraphics[width=0.9\linewidth]{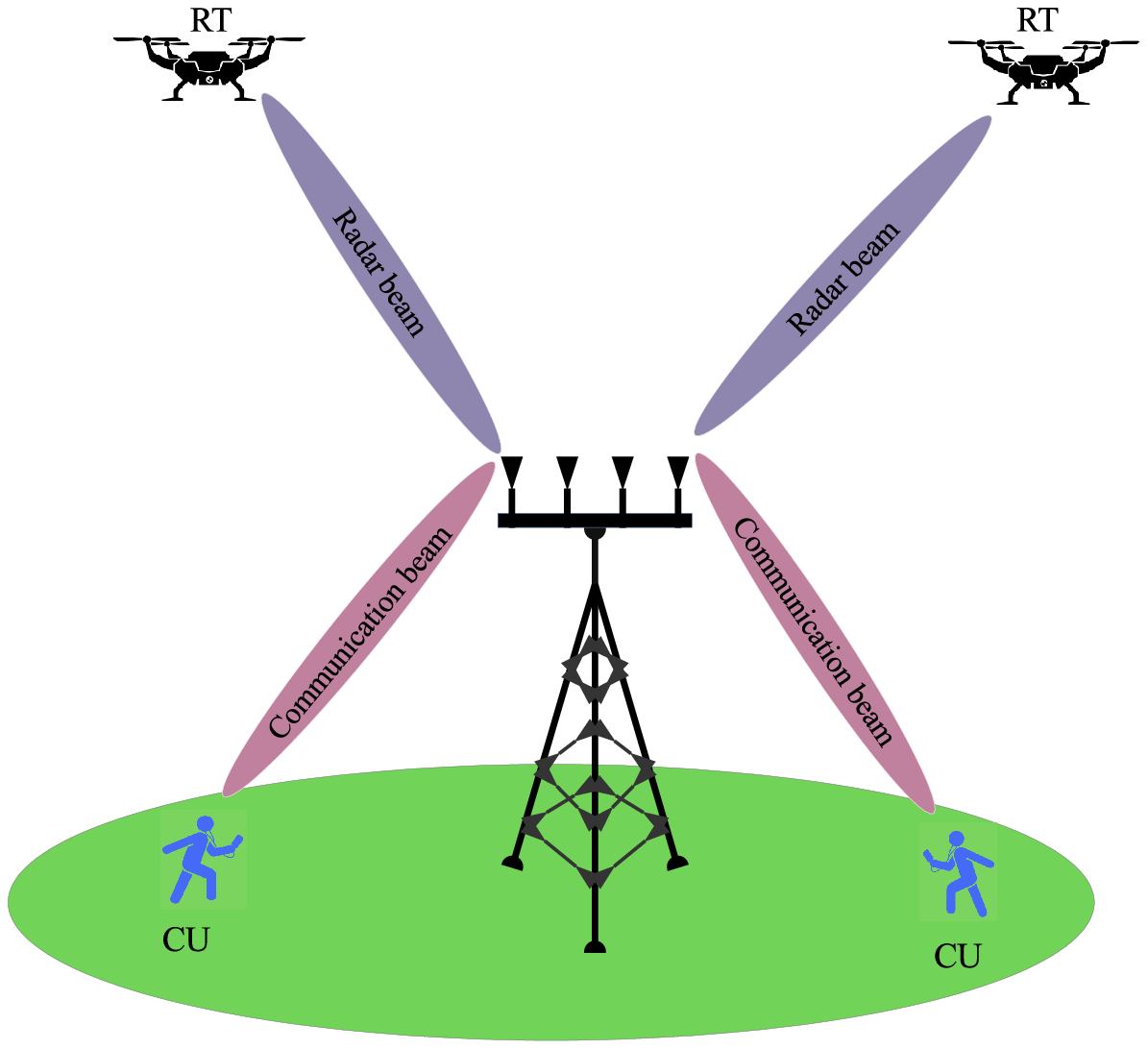}
    \caption{Illustration of mmWave MIMO ISAC system.}
    \label{fig:sys_1}
\end{figure}

\begin{figure}
    \centering
    \vspace{-1cm}
    \includegraphics[width=\linewidth]{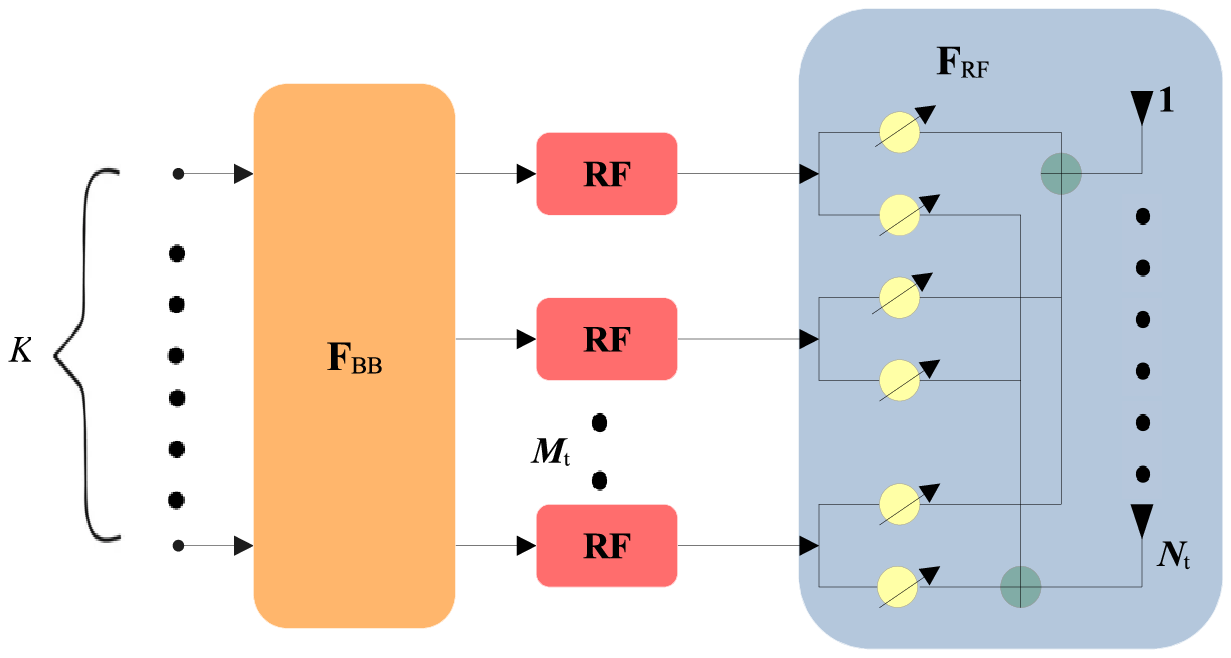}
    \vspace{-1.8cm}
    \caption{Illustration of the hybrid TPC structure at ISAC BS.}
    \label{fig:sys_2}
\end{figure}

%\begin{figure*}[t]%
%\centering
%\begin{subfigure}{\columnwidth}
%\hspace{1cm}
%\vspace{1cm}
%\includegraphics[width=0.8\columnwidth]{BASE_STATION.eps}%
%\vspace{-3.5cm}
%\caption{}
%\label{figure:Fig1}
%\end{subfigure}\hfill%
%\begin{subfigure}{\columnwidth}
%\hspace{-1cm}
%\vspace{3cm}
%\includegraphics[width=1.1\columnwidth]{Interior_BS.eps}
%\vspace{-5cm}
%\caption{}
%\label{figure:Fig2}
%\label{fig:system_model}
%\caption{(a) Illustration of mmWave MIMO ISAC system; (b)Illustration of the hybrid TPC structure at ISAC BS.}
%\vspace{-6mm}
%\end{figure*}
%\begin{figure}[t]
%\begin{center}
%\includegraphics [width=8cm]{BASE_STATION.eps}
%\vspace{-0.5cm}
%\caption{Illustration of mmWave MIMO ISAC system.}
%\label{figure:Fig1}
%\end{center}
%\vspace{-2cm}
%\end{figure}
%\begin{figure}[h!]
%\centering
%\hspace{-1.7cm}
%\includegraphics [width=9.5cm]{Interior_BS.eps}
%\vspace{-1.5cm}
%\caption{Illustration of the hybrid TPC structure at ISAC BS.}
%\label{figure:Fig2}
%\vspace{-5mm}
%\end{figure}
As shown in Fig. \ref{fig:sys_1}, we consider a mmWave MIMO ISAC system, in which an ISAC BS transmits $K=M+L$ data streams, while it serves $M$ communication users (CUs) and aims for detecting $L$ radar targets (RTs) in the scattering scene. The ISAC BS is equipped with $N_\mathrm{t}$ transmit antennas/ receive antennas, whereas each CU is equipped with $N_\mathrm{r}$ antennas. A fully connected hybrid architecture is exploited at the ISAC BS with only $M_\mathrm{t}$ RF chains as shown in Fig. \ref{fig:sys_2}, where we have $K\leq M_\mathrm{t}\leq N_\mathrm{t}$ to reduce the cost and power consumption. Furthermore, each CU is assumed to have a single RF chain, which is connected to each antenna via $N_\mathrm{r}$ phase shifters. Therefore, fully analog beamforming is used at each CU. Notably, for serving $M$ independent CUs and sensing $L$ different RTs simultaneously, $M_\mathrm{t}\triangleq K$ number of RF chains are required at the ISAC BS to form $M$ different beams for the CUs and $L$ beams for the RTs. Let us define the transmit signal $\mathbf{s}\in \mathbb{C}^{K \times 1}$ 
as
\begin{equation}
 \mathbf{s} = \begin{bmatrix}
           \mathbf{x}_1 \\
           \mathbf{x}_2 \\
         \end{bmatrix},
         %\vspace{-1mm}
\end{equation}
where $\mathbf{x}_1=[s_1, s_2,\hdots, s_{M}]^T \in \mathbb{C}^{M\times 1}$ and $\mathbf{x}_2=[s_{M+1},\hdots, s_{K}]^T\in {\mathbb C}^{L \times 1}$ are statistically independent and the signal $\mathbf{s}$ has a zero mean, i.e., it satisfies $\mathbb{E}\{\mathbf{s}\}=\mathbf{0}$ and has a covariance matrix given by $\mathbf{E}\{\mathbf{s}\mathbf{s}^H\}=\frac{P_\mathrm{t}}{K}\mathbf{I}_{K}$.
Note that while both $\mathbf{x}_1$ and $\mathbf{x}_2$ are utilized for radar target detection, only $\mathbf{x}_1$ is used for downlink communication, whereas $\mathbf{x}_2$ carries no useful information.
%is intended for the CUs and $\mathbf{x}_2=[s_{M+1},\hdots, s_{K}]^T\in {\mathbb C}^{L \times 1}$ is used for the detection of the $L$ RTs. Furthermore, we assume that both the signals meant for the CUs and the RTs are statistically independent with zero mean, i.e., satisfy $\mathbb{E}\{\mathbf{s}\}=\mathbf{0}$ and $\mathbf{E}\{\mathbf{s}\mathbf{s}^H\}=\frac{P_\mathrm{t}}{K}\mathbf{I}_{K}$.
Following the hybrid architecture \cite{HBF_7}, the transmitted signal $\mathbf{s}$ is first precoded by the baseband transmit precoder (TPC) of $\mathbf{F}_\mathrm{BB}=[\mathbf{f}_{{\rm BB},1}, \hdots,\mathbf{f}_{{\rm BB},M_\mathrm{t}}]\in {\mathbb C}^{{M_\mathrm{t}} \times K}$, followed by the RF TPC of $\mathbf{F}_\mathrm{RF}\in {\mathbb C}^{{N_\mathrm{t}} \times {M_\mathrm{t}}}$.  Upon assuming the availability of full CSI at each CU, the received signal $\mathbf {y}_{m}\in \mathbb{C}^{N_{\rm r}\times 1}$ of the $m$th CU is given as
\begin{subequations}\label{eqn:rx signal_1}
\begin{align}
&\mathbf {y}_{m}=\mathbf{ H}_m \mathbf{F}_{\rm RF}\mathbf{F}_{\rm BB}\mathbf{s} + \mathbf{n}_{m}, 
\\
&=\mathbf{ H}_m \mathbf{F}_{\rm RF}\mathbf{f}_{{\rm BB},m}s_m +
\sum_{n=1, n \neq m}^K\hspace{-0.3cm}\mathbf {H}_{m} \mathbf{F}_{\rm RF}\mathbf{f}_{{\rm BB},n} s_n+  \mathbf{n}_{m}, 
\end{align}
\end{subequations}
%\begin{figure}[t]
%\setkeys{Gin}{width=\linewidth}
 %  \hspace{-5mm}
 %   \begin{subfigure}[t]{0.25\textwidth}
  %  \vspace{-0.1cm}
  %  \includegraphics[width=0.9\textwidth]{BASE_STATION.eps}
  % % \vspace{-2cm}
  % %\hspace{-10cm}
  %  \caption{} \label{fig:sys_1}
%\end{subfigure}
% %\hspace{5mm}
%\begin{subfigure}[t]{0.25\textwidth}
%\vspace{-0.8cm}
%\hspace{-1.5cm}
 %   \includegraphics[width=1.3\textwidth]{Interior_BS.eps}
 %   \vspace{-1.2cm}
 %   \caption{} \label{fig:sys_2}
%\end{subfigure}
%%\vspace{-1.7cm}
%\caption{(a) Illustration of mmWave MIMO ISAC system. (b) Illustration of the hybrid TPC structure at ISAC BS.}
%\vspace{-0.5cm}
%\end{figure}
%\vspace{-2cm}

%\begin{figure}[t]
        %\begin{subfigure}[b]{0.4\textwidth}
        %\hspace{.8cm}
          %      \includegraphics[width=0.9\linewidth]{BASE_STATION.eps}
           %      \caption{Illustration of mmWave MIMO ISAC system.}
            %    \label{fig:sys_1}
             %   \vspace{-5mm}
             %   \end{figure}
        %\end{subfigure}\hfill
        %\begin{figure}[t]
        %\vspace{-1cm}
        %\hspace{-1cm}
            %    \includegraphics[width=1\linewidth]{Interior_BS.eps}
             %  \vspace{-1.5cm}
              %  \caption{Illustration of the hybrid TPC structure at ISAC BS.}
               % \label{fig:sys_2}
               % \vspace{-5mm}
        %\end{figure}%
where $\mathbf{H}_m$ represents the narrowband block-fading mmWave MIMO channel between the ISAC BS and the $m$th CU, while $\mathbf{n}_{m}\sim\mathcal{CN}(\mathbf{0}, \sigma^2\mathbf{I})$ is the complex circularly symmetric additive white Gaussian noise (AWGN). Using the fully analog combiner $\mathbf{w}_m\in \mathbb{C}^{N_{\rm r} \times 1}$ at each CU, the processed received signal $\widetilde{y}_{m}$ at the $m$th CU is given by
\begin{equation}\label{eqn:prc rx signal_1}
\begin{aligned}
\hspace{-1mm}\widetilde{y}_{m}=  
  & \mathbf{w}_m^H\mathbf{H}_{m} \mathbf{F}_{\rm RF}\mathbf{f}_{{\rm BB},m}s_{m} +\hspace{-4mm}\sum_{n=1, n \neq m}^K\hspace{-4mm}\mathbf{w}_m^H  
  \mathbf{H}_{m} \mathbf{F}_{\rm RF}\mathbf{f}_{{\rm BB},n}s_n\hspace{-1mm}+\mathbf{w}_m^H\mathbf{n}_{m}.
 \end{aligned}
\end{equation}
As a result, the achievable rate of the $m$th CU is given by
%\vspace{-2mm}
\begin{equation}\label{eqn:R_m}
R_m\left(\mathbf{w}_{m}, \mathbf{F}_\mathrm{RF}, \mathbf{F}_\mathrm{BB}\right) = \log_{2}\left(1+\gamma_m\right),
%\vspace{-2mm}
\end{equation}

where $\mathbf{\gamma}_m$ is the signal power to MUI and noise power ratio (SINR) of the $m$th CU, which is given by
\begin{equation}\label{eqn:8}
\mathbf{\gamma}_m = \frac{\left\vert\mathbf{w}_m^H\mathbf{H}_m \mathbf{F}_{\rm RF}\mathbf{f}_{{\rm BB},m}\right\vert^2}
{\sum_{n=1, n \neq m}^{K}{\left\vert\mathbf{w}_m^H\mathbf{H}_m \mathbf{F}_{\rm RF}\mathbf{f}_{{\rm BB},n}\right\vert^2} + \sigma^2 \left\vert\left\vert\mathbf{w}_m\right\vert\right\vert^2_F}.
\end{equation}
%Therefore, the achievable sum-SE ${\cal R}_{\rm sum}$ and GM-SE ${\cal R}_{\rm GM}$ of the system under consideration are given by
%\begin{subequations}
%\begin{align}
%&\mathcal{R}_{\rm sum} = \sum_{m=1}^{M}R_m\left(\mathbf{w}_{m}, \mathbf{F}_\mathrm{RF}, \mathbf{F}_\mathrm{BB}\right),\label{eqn:R_sum} \\
%&\mathcal{R}_{\rm GM} = \left ({\prod_{m=1}^{M} R_{m}\left(\mathbf{w}_{m}, \mathbf{F}_\mathrm{RF}, \mathbf{F}_\mathrm{BB}\right)}\right)^{1/M}. \label{eqn:R_GM}
%\end{align}
%\end{subequations}}
%where $\mathbf{\gamma}_m$ is the signal power to MUI and noise power ratio (SINR) of the $m$th CU, which is given by
%\begin{equation}\label{eqn:8}
%\mathbf{\gamma}_m = \frac{\left\vert\mathbf{w}_m^H\mathbf{H}_m \mathbf{F}_{\rm RF}\mathbf{f}_{{\rm BB},m}\right\vert^2}
%{\sum_{n=1, n \neq m}^{K}{\left\vert\mathbf{w}_m^H\mathbf{H}_m \mathbf{F}_{\rm RF}\mathbf{f}_{{\rm BB},n}\right\vert^2} + \sigma^2 \left\vert\left\vert\mathbf{w}_m\right\vert\right\vert^2_F}.
%\end{equation}
\subsection{Channel model}
We employ the widely used Saleh-Valenzuela channel model of \cite{HBF_6} to characterize the wireless channel. This model captures the parameters of the multipath components for accurately characterizing the wireless channel, which includes the complex path losses, delays, angles-of-arrival (AoAs), and angles-of-departure (AoDs). The frequency-flat mmWave MIMO channel for communication between ISAC BS and $m$th CU, equipped with uniform linear arrays (ULAs), is expressed as \cite{ISAC_mmWave_10,ISAC_mmWave_35}
\begin{equation}\label{eqn:channel}
\mathbf{H}_{m}= \sum_{i=1}^{N^{\rm p}_m}\alpha_{m,i}\mathbf{a}_{\rm r}(\phi_{m,i})\mathbf{a}_{\rm t}^H(\theta_{m,i}), 
\end{equation}
where $N^p_m$ denotes the number of multipath components\footnote{It is important to note that some RTs may contribute to increasing the number of multipath components in the communication channel. However, this paper neglects this effect due to the high path loss and the strong susceptibility of millimeter waves to blockages \cite{ISAC_mmWave_1,ISAC_mmWave_3,ISAC_mmWave_1_1,ISAC_mmWave_12_1,ISAC_mmWave_15,ISAC_mmWave_4,ISAC_mmWave_5,ISAC_mmWave_6,ISAC_mmWave_14,ISAC_mmWave_13,ISAC_mmWave_12,ISAC_mmWave_IRS_1,ISAC_mmWave_IRS_2,ISAC_mmWave_IRS_3,new_13,new_12}.} in $\mathbf{H}_m$. The quantity $\alpha_{m,i}$ in (\ref{eqn:channel}) is the channel gain of the $i$th multipath that is distributed as $\mathcal{CN}(0,\gamma_m^210^{-0.1PL(d_m)}), \forall l=\{1,\hdots, N^p_m\}$, where $\gamma_m=\sqrt{N_\mathrm{r}N_\mathrm{t}/N^p_m}$ denotes the normalization factor associated with $PL(d_m)$ as the path loss that depends on the distance $d_m$ associated with the corresponding link \cite{ISAC_mmWave_10,ISAC_mmWave_35}. Moreover, the coefficients $\alpha_{m,i}$ are independently distributed for different $i$.
Furthermore, $\mathbf{a}_{\rm t}(\theta_{m,i})\in \mathbb{C}^{N_\mathrm{t}\times 1}$ and 
$\mathbf{a}_{\rm r}(\phi_{m,i})\in \mathbb{C}^{N_\mathrm{r}\times 1}$ are the transmit and 
receive array response vectors for the channel matrix $\mathbf{H}_m$, associated with the AoAs 
$\phi_{m,i}$ and AoDs $\theta_{m,i}$, respectively. The array response vectors for the ULAs at the ISAC BS and each CU are given by
\begin{equation}\label{eqn:array respo}
\begin{aligned} 
\mathbf{a}_{z}\left (\delta \right)=\frac {1}{\sqrt{N_z}}\biggl [1, \ldots, e^{j \frac {2 \pi }{\lambda } d \left(n \cos \delta\right)} ,\ldots, e^{j \frac {2 \pi }{\lambda } d \left ({N_z-1)\cos \delta}\right)}\biggr]^{T},
\end{aligned}
\end{equation}
where $z\in \{\rm r,t\}$ and $\delta\in \{\phi, \theta\}$. The quantities $\lambda$ and $d$ represent the wavelength and the antenna spacing, where the latter is assumed to be equal to half of the wavelength. In addition, $0\leq n\leq N_z$ denote the indices of the elements of the ULA of size $N_z$ in the 1D plane. 
%We define the transmit array response matrix $\mathbf{A}^{\rm t}_m\in \mathbb{C}^{N^{\rm p}_m\times N_\mathrm{t}}$, the complex gain matrix $\mathbf{G}_m\in \mathbb{C}^{N^{\rm p}_m\times N^{\rm p}_m}$ and the receive array response matrix $\mathbf{A}^{\rm r}_m\in \mathbb{C}^{N_\mathrm{r}\times N^{\rm p}_m}$ for the channel $\mathbf{H}_m$ by stacking the antenna array response vectors and complex gains of the channel as
%\begin{equation}\label{eqn:array respo matrix_1}
%\mathbf {A}^{\rm t}_{m} = \Big[\mathbf{a}_{\rm t}(\theta_{m,0}) \hdots \mathbf{a}_{\rm t}(\theta_{m,N^{\rm p}_m})\Big],
%\end{equation}
%\begin{equation}\label{eqn:array respo matrix_2} 
%\mathbf{G}_m = \mathcal{D}\Big(\left[\alpha_{m,0},\hdots,\alpha_{m,N^{\rm p}_m}\right]\Big),
%\end{equation}
%\begin{equation}\label{eqn:array respo matrix_3}
%\mathbf {A}^{\rm r}_m = \Big[\mathbf{a}_{\rm r}(\phi_{m,0}) \hdots \mathbf{a}_{\rm r}(\phi_{m,N^{\rm p}_m})\Big].
%\end{equation}
%Hence, the mmWave MIMO channel $\mathbf{H}_m$ can be compactly expressed as 
%\begin{equation}\label{eqn:deco_channel}
%\begin{aligned} 
%\mathbf{H}_m = \mathbf {A}^{\rm r}_m\mathbf{G}_m(\mathbf{A}^{\rm t}_{m})^H.
%\end{aligned}
%\end{equation}
\subsection{Radar model}\label{Radar Model}
At the ISAC BS, we assume that the same antenna array is used for transmitting and receiving radar signals \cite{ISAC_MIMO_6}. However, this leads to a signal leakage problem in such an architecture, which can be efficiently mitigated via RF and baseband cancellers, as discussed in \cite{application_3}. Moreover, there is a signal-dependent interference due to the presence of multiple targets in the system, and the noise term includes self-interference. Thus, the signal received for an RT located at an angle of $\theta_l$ at the ISAC BS can be written as
%\vspace{-2mm}
\begin{equation} \label{eqn:radar_rec}
\begin{aligned}
\mathbf {y}_{rad}(\theta_l)=&\tau_l\mathbf{a}_\mathrm{t}(\theta_l)\mathbf{a}_\mathrm{t}^{T}(\theta _l)\mathbf{F}_{\rm RF}\mathbf{F}_{\rm BB}\mathbf{s} \\
&+ \sum_{j=1, j\neq l}^{L}\tau_j\mathbf{a}_\mathrm{t}(\theta_j)\mathbf{a}_\mathrm{t}^{T}(\theta _{j})\mathbf{F}_{\rm RF}\mathbf{F}_{\rm BB}\mathbf{s}+\mathbf{n}_\mathrm{r}, 
\end{aligned}
%\vspace{-2mm}
\end{equation}
where $\tau_l$ and $\tau_j$ are the reflection coefficient for the desired target and $j$th interference, and $\mathbf{n}_\mathrm{r}\sim \mathcal{CN}(\mathbf{0}, \mathbf{I})$ is the complex circularly symmetric AWGN noise. 
In particular, we consider a sensing scenario in which the ISAC BS does have a-prior knowledge about the multiple RTs \footnote{Such scenario corresponds to the target tracking phases, in which the ISAC BS needs to track the $L$ RTs by estimating their %coefficients/
angles, with their initial parameters known in advance \cite{ISAC_mmWave_1_1,ISAC_mmWave_12}.
} such that the ISAC BS aims to estimate the RT’s %coefficients $\{\tau_l\}_{l=1}^L$ and 
angles $\{\theta_l\}_{l=1}^L$.
%where $\mathbf{r}_t, \mathbf{r}_i$ and $\mathbf {n}_{r}$ denote the desired target, interference, and noise signals in the radar sensing environment, respectively. 
%Specifically, the signal $\mathbf{r}_t$ and $\mathbf{r}_i$ are modeled as
%\begin{subequations}
%\begin{align}
%&\mathbf{r}_{t}=\mathbf{a}_\mathrm{t}(\theta_l)\mathbf{a}_\mathrm{t}^{T}(\theta_l)\mathbf{F}_{\rm RF}\mathbf{F}_{\rm BB}\mathbf{s},\\
%&\mathbf{r}_{i}=\sum_{j\neq l}^{L}\tau_j\mathbf{a}_\mathrm{t}(\theta_j)\mathbf{a}_\mathrm{t}^{T}(\theta _{j})\mathbf{F}_{\rm RF}\mathbf{F}_{\rm BB}\mathbf{s},
%\end{align}
%\end{subequations} 
Based on the received signal (\ref{eqn:radar_rec}), one can estimate the interest angles of the RTs by employing the well-known multiple signal classification (MUSIC) algorithm \cite{ISAC_mmWave_1_1}. In order to detect multiple RTs, the ISAC BS scans different angles of the space and maximizes the beampattern gains towards the potential directions of interest. 
It is important to highlight that, in this formulation, we do not require the full CSI (e.g., Doppler shifts, path delays, or complex path gains) of the radar targets, and the angular information alone is sufficient for obtaining the desired transmit beampattern for the sensing functionality.
Furthermore, the transmit beampattern gain towards the RT located at $\theta_l$ is given by
\begin{equation}\label{eqn:tx beampattern}
G(\theta_l, \mathbf{F}_\mathrm{RF}, \mathbf{F}_\mathrm{BB}) = \mathbf{a}_\mathrm{t}^H(\theta_l) \mathbf{F}_\mathrm{RF}\mathbf{F}_\mathrm{BB}\mathbf{F}_\mathrm{BB}^H\mathbf{F}_\mathrm{RF}^H\mathbf{a}_\mathrm{t}(\theta_l).
\end{equation}
Hence, one needs to maximize the transmit beampattern gain towards the RTs for maximizing their probability of detection \footnote{In target tracking, once the approximate directions of RTs are known, maximizing the transmit beampattern gain toward these directions enhances detection and improves tracking accuracy \cite{ISAC_MIMO_6,ISAC_MIMO_8,ISAC_MIMO_7,ISAC_MIMO_16,ISAC_MIMO_17}.}.
Let us assume that $\{G_{\mathrm{d}}(\theta _{l})\}_{l=1}^L$ is the desired beampattern at the $l$th sampled angle, which specifies the desired transmit power distribution at the $L$ angles $\{\theta_l\}_{l=1}^L$ in the space \cite{ISAC_MIMO_6}. Therefore, to evaluate the sensing performance, one must compare the transmit beampattern of the ISAC BS designed to the ideal one. Moreover, we consider the MSE between the desired and designed transmit beampatterns to evaluate the RBPS, which is formulated as 
%\vspace{-2mm}
\begin{equation} \label{eqn:beampattern simil}
\begin{aligned}
\Psi(\beta,\mathbf{F}_\mathrm{RF}, \mathbf{F}_\mathrm{BB})&\triangleq \frac{1}{L}\sum _{l=1}^{L}\left|\beta G_{\mathrm{d}}(\theta_{l})-G(\theta_l, \mathbf{F}_\mathrm{RF}, \mathbf{F}_\mathrm{BB}) \right|^{2},
\end{aligned}
%\vspace{-2mm}
\end{equation}
where $\beta$ is the scaling factor, which is used to regulate the scaling level of $G_{\mathrm{d}}(\theta_{l})$ such that the transmit beampattern can better match the scaled version of the desired beampattern. 
It is worth noting that the MSE as a measure of RBPS is particularly suitable for mmWave-based ISAC systems because of their reliance on highly directional beams, sparse channel properties, large antenna arrays, and the need for accurate dual-functional operation.
Therefore, minimizing the MSE $\Psi(\beta,\mathbf{F}_\mathrm{RF}, \mathbf{F}_\mathrm{BB})$ is essential for enhancing the detection performance of the RTs in such systems.

\section{sum-SE Maximization}\label{problem formulation}
This section seeks to design the hybrid TPC $\mathbf{F}_\mathrm{RF}, \mathbf{F}_\mathrm{BB}$ and RCs  $\mathbf{w}_m, \forall m$, that maximize the sum-SE of the CUs, considering the RBPS, the total transmit power and constant modulus constraints of the elements of $\mathbf{F}_\mathrm{RF}$ and $\mathbf{w}_m$. As a result, the pertinent optimization problem is given by
\begin{subequations}\label{eqn:system optimization}
\begin{align}
\mathcal{P}_{0}: \hspace{8mm} &\mathop{\max}_{\bigl\{\mathbf{w}_{m}\bigr\}_{m=1}^M,\mathbf{F}_\mathrm{RF},\mathbf{F}_\mathrm{BB}} \sum_{m=1}^{M}R_m\left(\mathbf{w}_{m}, \mathbf{F}_\mathrm{RF}, \mathbf{F}_\mathrm{BB}\right) \\
&\text {s.t.} \quad \Psi(\beta,\mathbf{F}_\mathrm{RF}, \mathbf{F}_\mathrm{BB})  \leq \epsilon, \label{constr:RBPS_1} \\
&\quad\quad\left\vert\mathbf{F}_\mathrm{RF}(i,j)\right\vert = \frac{1}{\sqrt{N_\mathrm{t}}}, \forall i, j, \label{constr:RF_1}\\
&\quad\quad\left\vert\mathbf{w}_m(i)\right\vert = \frac{1}{\sqrt{N_\mathrm{r}}}, \forall i, m\label{constr:comb_1}\\
&\quad\quad\|\mathbf{F}_\mathrm{RF}\mathbf{F}_\mathrm{BB}\|_F^2  \leq P_\mathrm{t}, \label{constr:TP_1}
\end{align}
\end{subequations}
where $\epsilon$ denotes the maximum MSE tolerance of the RBPS, which regulates the detection performance of the RTs. The second and third constraints in the above optimization problem are the constant modulus constraints imposed due to the phase shifters of the RF TPC and RC, whereas $P_\mathrm{t}$ is the transmit power constraint at the ISAC BS. 
The optimization problem $\mathcal{P}_0$ is highly non-convex due to the non-convex nature of the objective function and, owing to the highly non-convex quadratic RBPS constraint and non-convex constant modulus constraint, which renders the problem difficult to solve. Moreover, the variables $\bigr\{\mathbf{w}_{m}\bigr\}_{m=1}^M,\mathbf{F}_\mathrm{RF},\mathbf{F}_\mathrm{BB}$ are coupled in both the objective function and the constraints, which makes the problem even more challenging. To solve this demanding problem, we propose an approach that sequentially updates one block of variables, while fixing all others at a time. Initially, we decouple the optimization problem (\ref{eqn:system optimization}) into two sub-problems, where the first sub-problem focuses on the design of the hybrid TPC components $\mathbf{F}_\mathrm{RF},\mathbf{F}_\mathrm{BB}$ for a fixed RC $\mathbf{w}_{m}$, while the second sub-problem seeks to design the RCs $\bigr\{\mathbf{w}_{m}\bigr\}_{m=1}^M$ for the TPCs obtained above. We discuss the proposed sequential method in the subsequent subsections.
\subsection{Hybrid TPC design for fixed RC}\label{hybrid precoder and optimal power allocation}
In this subsection, we focus on the design of the hybrid TPC $\mathbf{F}_\mathrm{RF}, \mathbf{F}_\mathrm{BB}$ for the fixed RCs $\mathbf{w}_m, \forall m$, which maximizes the sum-SE of the CUs while meeting the RBPS constraint of the RTs. For the fixed combiner $\mathbf{w}_m$, the first sub-problem of (\ref{eqn:system optimization}) is transformed to
\begin{subequations}\label{eqn:TPC optimization_1}
\begin{align}
&\mathop{\max}_{\mathbf{F}_\mathrm{RF},\mathbf{F}_\mathrm{BB}}\sum_{m=1}^{M}R_m\left( \mathbf{F}_\mathrm{RF}, \mathbf{F}_\mathrm{BB}\right) \\ 
& \text {s.t.} \quad \text{(\ref{constr:RBPS_1}), (\ref{constr:RF_1}), (\ref{constr:TP_1})}.
\end{align}
\end{subequations}
The above optimization problem is non-convex due the non-convex nature of the objective function, arising as a consequence of the fractional term in the $\mathrm{log}$ function and the non-convex constraints. To tackle the non-convexity, we propose a two-stage hybrid TPC design, in which the first-stage maximizes the sum-SE of the CUs, while the MUI is mitigated in the second-stage. With this in mind, we write the baseband TPC as 
\begin{equation}\label{eqn:BB}
\begin{aligned}
\mathbf{F}_\mathrm{BB} = \left[\mathbf{F}^\mathrm{C}_\mathrm{BB} \hspace{2mm}\mathbf{F}^\mathrm{R}_\mathrm{BB}\right],
\end{aligned}
\end{equation}
where $\mathbf{F}^\mathrm{C}_\mathrm{BB}\in \mathbb{C}^{M_\mathrm{t}\times M}$ is meant for the CUs, while $\mathbf{F}^\mathrm{R}_\mathrm{BB}\in \mathbb{C}^{M_\mathrm{t}\times L}$ is used for the RTs. We now decompose $\mathbf{F}^\mathrm{C}_\mathrm{BB}$ as $\mathbf{F}^\mathrm{C}_\mathrm{BB}=\widetilde{\mathbf{F}}^{\mathrm{C}, 1}_\mathrm{BB}\widetilde{\mathbf{F}}^{\mathrm{C}, 2}_\mathrm{BB}$, where $\widetilde{\mathbf{F}}^{\mathrm{C}, 1}_\mathrm{BB}=\left[\widetilde{\mathbf{f}}^{\mathrm{C}, 1}_{{\rm BB},1}, \hdots, \widetilde{\mathbf{f}}^{\mathrm{C}, 1}_{{\rm BB},M}\right]\in \mathbb{C}^{M_\mathrm{t}\times M}$ maximizes the sum-SE, while $\widetilde{\mathbf{F}}^{\mathrm{C}, 2}_\mathrm{BB}=\left[\widetilde{\mathbf{f}}^{\mathrm{C}, 2}_{{\rm BB},1}, \hdots, \widetilde{\mathbf{f}}^{\mathrm{C}, 2}_{{\rm BB},M}\right]\in \mathbb{C}^{M\times M}$ mitigates the MUI emanating from the CUs. Therefore, in the first-stage, we jointly design $\mathbf{F}_\mathrm{RF}$ and $\widetilde{\mathbf{F}}^{\mathrm{C}, 1}_\mathrm{BB}$ for maximizing the sum-SE of the CUs, while ignoring the MUI. Subsequently, based on the effective baseband channel, $\widetilde{\mathbf{F}}^{\mathrm{C}, 2}_\mathrm{BB}$ and $\mathbf{F}^\mathrm{R}_\mathrm{BB}$ are designed in the second-stage for mitigating the MUI arriving from the other CUs and RTs. 
Based on the above decomposition, the received signal (\ref{eqn:prc rx signal_1}) of the $m$th CU can be rewritten as 
\begin{subequations}\label{eqn:rx signal_2}
\begin{align}
\widetilde{y}_{m}=&\mathbf{w}^H_m\mathbf{H}_m \mathbf{F}_{\rm RF}\left[\widetilde{\mathbf{F}}^{\mathrm{C}, 1}_\mathrm{BB}\widetilde{\mathbf{F}}^{\mathrm{C}, 2}_\mathrm{BB}, \mathbf{F}^\mathrm{R}_\mathrm{BB}\right]\mathbf{s} + \mathbf{w}^H_m\mathbf{n}_{m}, 
\\=&\mathbf{w}^H_m\mathbf{H}_m \mathbf{F}_{\rm RF}\widetilde{\mathbf{F}}^{\mathrm{C}, 1}_\mathrm{BB}\widetilde{\mathbf{F}}^{\mathrm{C}, 2}_\mathrm{BB}\mathbf{x}_1 +\mathbf{w}^H_m\mathbf{H}_m \mathbf{F}_{\rm RF}\mathbf{F}^\mathrm{R}_\mathrm{BB}\mathbf{x}_2 \notag\\
&+ \mathbf{w}^H_m\mathbf{n}_{m}.
\end{align}
\end{subequations}
As seen from $(\ref{eqn:rx signal_2})$, the first and second quantities are meant for the CUs and RTs, respectively. Therefore, the optimization problem (\ref{eqn:TPC optimization_1}) can be recast as 
\begin{subequations}\label{eqn:hybrid TPC 1}
\begin{align}
&\mathop{\max}_{\mathbf{F}_\mathrm{RF}, \widetilde{\mathbf{F}}^{\mathrm{C}, 1}_\mathrm{BB}, \widetilde{\mathbf{F}}^{\mathrm{C}, 2}_\mathrm{BB}, \mathbf{F}^\mathrm{R}_\mathrm{BB}}\sum_{m=1}^{M}R_m\left( \mathbf{F}_\mathrm{RF}, \mathbf{F}_\mathrm{BB}\right)\\
&\text {s.t.} \quad \Psi\left(\beta,\mathbf{F}_\mathrm{RF},  \left[\widetilde{\mathbf{F}}^{\mathrm{C}, 1}_\mathrm{BB}\widetilde{\mathbf{F}}^{\mathrm{C}, 2}_\mathrm{BB} \hspace{2mm}\mathbf{F}^\mathrm{R}_\mathrm{BB}\right]\right) \leq \epsilon, \label{constr:RBPS_2}\\
&\quad\quad\left\vert\mathbf{F}_\mathrm{RF}(i,j)\right\vert = \frac{1}{\sqrt{N_\mathrm{t}}}, \forall i, j, \label{constr:RF_2} \\
&\quad\quad\left\vert\left\vert\mathbf{F}_\mathrm{RF}\left[\widetilde{\mathbf{F}}^{\mathrm{C}, 1}_\mathrm{BB}\widetilde{\mathbf{F}}^{\mathrm{C}, 2}_\mathrm{BB} \hspace{2mm}\mathbf{F}^\mathrm{R}_\mathrm{BB}\right]\right\vert\right\vert_F^2\leq P_\mathrm{t}. \label{constr:TP_2}
\end{align}
\end{subequations}
Assuming that the MUI is cancelled in the second-stage of the TPC design, the mutual information of the $m$th CU, while ignoring the MUI, is defined as
\begin{equation}\label{eqn:mutual information_1}
\begin{aligned} 
\mathcal{I}_m=\log_2\left(1+\mathbf{w}_m^H\mathbf{H}_m \mathbf{F}_{\rm RF}\widetilde{\mathbf{f}}^{\mathrm{C}, 1}_{{\rm BB},m}\left(\widetilde{\mathbf{f}}^{\mathrm{C}, 1}_{{\rm BB},m}\right)^H\mathbf{F}^H_{\rm RF}\mathbf{H}^H_m\mathbf{w}_m\right).
\end{aligned}
\end{equation}
Since the BS and each CU design their TPC and RC separately for maximizing the mutual information, the TPC design problem in the first-stage can be decoupled into multiple distinct single-CU TPC design problems. Let us define the singular value decomposition (SVD) of $\mathbf{H}_m$ as
\begin{equation}\label{eqn:SVD} 
\mathbf{H}_{m} =\mathbf{U}_{m}\mathbf{\Sigma}_{m}\mathbf{V}^H_{m},
\end{equation}
where $\mathbf{U}_{m}\in\mathbb{C}^{N_{\rm r}\times N_{\rm r}}, \mathbf{V}_{m}\in\mathbb{C}^{N_{\rm t}\times N_{\rm t}}$ are the left and right singular matrices, while $\mathbf{\Sigma}_{m}\in\mathbb{C}^{N_{\rm r}\times N_{\rm t}}$ is the diagonal matrix containing the singular values. Hence, the optimal fully-digital TPC $\mathbf{f}^\mathrm{opt}_m$ of the $m$th CU is given by the specific vector from the right singular matrix $\mathbf{V}$ corresponding to the highest singular value, i.e., $\mathbf{f}^\mathrm{opt}_m=\mathbf{V}_{m}(1,:)$.
By considering the hybrid TPC of the $m$th CU as the optimal fully digital TPC, i.e., $\mathbf{F}_{\rm RF}\widetilde{\mathbf{f}}^{\mathrm{C}, 1}_{{\rm BB},m}=\mathbf{f}^\mathrm{opt}_m$, $\mathcal{I}_m$ can be approximated as 
\begin{subequations}\label{eqn:mutual information_2}
\begin{align} 
\mathcal{I}_m &\approx \log _2\left(1+{\mathbf{\Sigma}}^2_m(1,1)\Omega_m\right) \\ 
&=\Bigg [\log _2\left(1+\mathbf{\Sigma }^2_m(1,1)-\mathbf{\Sigma}^2_m(1,1)(1-\Omega _m)\right)\Bigg ] \\ 
&\approx \log_2\left(1+\mathbf{\Sigma}^2_m\left(1,1\right)\right)-\frac{\mathbf{\Sigma}^2_m\left(1,1\right)(1-\Omega _m)}{1+\mathbf{\Sigma}^2_m(1,1)} \\ 
&\approx \log _2 \left(1+\mathbf{\Sigma}^2_m(1,1) \right)-\left(1- \Omega_m \right),
\end{align}
\end{subequations}
where $\Omega_m = \|(\mathbf{f}^\mathrm{opt})^H_m\mathbf{F}_{\rm RF}\widetilde{\mathbf{f}}^{\mathrm{C}, 1}_{{\rm BB},m}\|^2_F$. For maximizing $\mathcal{I}_m$, one has to minimize the quantity $\left(1- \Omega_m \right)$, which is the average squared chordal distance between the optimal full-RF TPC $\mathbf{f}^\mathrm{opt}_m$ and the hybrid TPC $\mathbf{F}_{\rm RF}\widetilde{\mathbf{f}}^{\mathrm{C}, 1}_{{\rm BB},m}$. As a result, the hybrid TPC design problem in the first-stage can be reformulated as 
\begin{equation}\label{eqn:hybrid TPC 3}
\begin{aligned}
\mathcal{P}_1: \hspace{2mm}&\mathop{\min}_{\mathbf{F}_\mathrm{RF}, \widetilde{\mathbf{F}}^{\mathrm{C}, 1}_\mathrm{BB}}\|\mathbf{F}^\mathrm{opt}-\mathbf{F}_\mathrm{RF}\widetilde{\mathbf{F}}^{\mathrm{C}, 1}_\mathrm{BB}\|^2_F \\
&\text {s.~t.} ~ \text{(\ref{constr:RBPS_2}), (\ref{constr:RF_2}), (\ref{constr:TP_2}),}
\end{aligned}
\end{equation}
where $\mathbf{F}^\mathrm{opt}=\left[\mathbf{f}^\mathrm{opt}_1, \hdots, \mathbf{f}^\mathrm{opt}_M\right]\in \mathbb{C}^{N_\mathrm{t}\times M}$. The optimization problem (\ref{eqn:hybrid TPC 3}) is still non-convex and difficult to solve due to the non-convex RBPS and constant modulus constraints, and owing to the tightly coupled variables in both the objective function and the constraints. To solve this problem, we adopt the alternating optimization technique in conjunction with the MM and the RCG techniques, where the RF and baseband TPCs are designed iteratively, until convergence is achieved. The proposed method is discussed next.
\subsubsection{Analog beamformer design}
For a fixed baseband TPC $\mathbf{F}_\mathrm{BB}$, the RF TPC design problem can be formulated as 
\begin{subequations}\label{eqn:RF_1}
\begin{align}
&\mathop{\min}_{\mathbf{F}_\mathrm{RF}}\|\mathbf{F}^\mathrm{opt}-\mathbf{F}_\mathrm{RF}\widetilde{\mathbf{F}}^{\mathrm{C}, 1}_\mathrm{BB}\|^2_F \\
&\text {s.t.} \quad
\Psi(\beta,\mathbf{F}_\mathrm{RF}) \leq \epsilon %\label{constr:RBPS_3}
~\text{and}~\left\vert\mathbf{F}_\mathrm{RF}(i,j)\right\vert = \frac{1}{\sqrt{N_\mathrm{t}}}, \forall i, j. %\label{constr:RF_3}
\end{align}
\end{subequations}

Due to the non-convex RBPS and constant modulus constraints, (\ref{eqn:RF_1}) is still non-convex. In order to solve this problem, we develop a novel Riemannian-majorization and minimization (MM) conjugate gradient (RMCG) algorithm to address the non-convex RBPS and the constant modulus constraints, which is discussed next. 

To deal with the non-convex quadratic MSE function $\Psi(\beta,\mathbf{F}_\mathrm{RF})$, we present a family of surrogate functions that can be easily obtained using the majorization-minimization (MM) technique. Toward this, let us determine the optimal value of the quantity $\beta$ in the RBPS constraint of $(\ref{eqn:RF_1})$. As $\Psi(\beta,\mathbf{F}_\mathrm{RF})$ is a quadratic-convex function, the optimal value of $\beta$ can be evaluated using the first-order optimal condition, which is given by
\begin{subequations}\label{eqn:beta}
\begin{align}
&\frac{\partial \Psi\left(\beta, \mathbf{F}_\mathrm{RF}\right)}{\partial \beta}=0 \\
&\beta ^{\star}=\frac{\sum _{l=1}^{L}G_{\mathrm{d}}(\theta_l) \mathbf{a}^H_l\mathbf{d}}{\sum _{l=1}^{L}G_{\mathrm{d}}^{2}(\theta_l)}, 
\end{align}
\end{subequations}
where $\mathbf{a}_l = \text{vec}(\mathbf{A}_l)$ with $\mathbf{A}_l$ as the transmit array response matrix at the $l$th sampled angle $( \theta_l)$, which is defined as $\mathbf{A}_l\triangleq \mathbf{a}_\mathrm{t}( \theta_l)\mathbf{a}^H_\mathrm{t}(\theta_l)$. Furthermore, the quantity $\mathbf{d}$ is defined as $\mathbf{d} = \text{vec}(\mathbf{F}_\mathrm{RF}\mathbf{F}_\mathrm{RF}^H)$.

Upon substituting the optimal $\beta ^{\star}$ into the function $\Psi\left(\beta, \mathbf{F}_\mathrm{RF}\right)$, and using some fundamental algebraic transformations, the RBPS constraint can be rewritten as
\begin{equation} \label{eqn:RF_2}
\begin{aligned}
\Psi(\mathbf{F}_\mathrm{RF})=\mathbf{d}^H \mathbf {C}\mathbf{d} \leq \epsilon, 
\end{aligned}
\end{equation}
where
\begin{align}
\mathbf {C}&\triangleq \frac{1}{L}\sum _{l=1}^{L}\mathbf {b}_{l}\mathbf {b}_{l}^{H}\label{eqn:RF_3_1},\\ 
\mathbf {b}_{l}&\triangleq \frac{G_{\mathrm{d}}(\theta_{l}) \sum _{l_{1}=1}^{L}G_{\mathrm{d}}(\theta_{l_1})\mathbf{a}_{l_1}}{\sum_{l_1=1}^{L}G_{\mathrm{d}}^{2}(\theta_{l_1})}-\mathbf{a}_l\label{eqn:RF_3}.  
\end{align}
Due to the quadratic and non-convex MSE function pertaining to the beampattern design in $(\ref{eqn:RF_2})$, we develop a set of surrogate functions for $\Psi(\mathbf{F}_\mathrm{RF})$ that are mathematically tractable for optimization. To achieve this, we leverage the MM technique, in which the surrogate function of $\Psi(\mathbf{F}_\mathrm{RF})$ at a current local point $\mathbf{F}^t_\mathrm{RF}$ in the $t$th iteration serves as an upper-bound, that can be designed as \cite{MM_1,MM_2}
\begin{equation}\label{eqn:RF_4}
\begin{aligned}
\Psi(\mathbf{F}_\mathrm{RF})& \leq \lambda\|\mathbf{d}\|^2_F + \Re \lbrace\mathbf{d}^H\mathbf {b}^{t}\rbrace + c_{1}^{t},  
\end{aligned}
\end{equation}
where $\lambda$ is the largest eigenvalue of the matrix $\mathbf{C}$. For brevity, the quantities $\mathbf{b}^{t}$ and $c_{1}^{t}$ are defined as
\begin{align}
\mathbf {b}^{t}&\triangleq 2(\mathbf {C}-\lambda\mathbf {I}) \mathbf{d}^t \label{eqn:RF_5},\\ 
c_{1}^{t}&\triangleq \left(\mathbf{d}^t\right)^H (\lambda\mathbf {I}-\mathbf {C}) \mathbf{d}^t\label{eqn:RF_5_1}. 
\end{align}
Since the baseband TPC $\mathbf{F}_\mathrm{BB}$ also affects the RBPS constraint, we follow the following approach to design the beampattern of the ISAC BS, which depends only on the RF TPC $\mathbf{F}_\mathrm{RF}$ by constraining $\mathbf{F}_\mathrm{BB}$ to be a unitary matrix. Considering a unitary baseband TPC and the total power constraint, the first term of the right hand side of (\ref{eqn:RF_4}) can be upper-bounded as \begin{equation}\label{eqn:RF_6}
\begin{aligned}
&\lambda\|\mathbf{d}\|_F^2 =\lambda \Big \Vert \sum _{j=1}^{M_\mathrm{t}}\mathbf{f}_{\mathrm{RF}, j}\mathbf{f}^H_{\mathrm{RF}, j} \Big \Vert _{F}^{4}\\ 
&\quad \leq \lambda\left(\sum _{j=1}^{M_\mathrm{t}}\Vert \mathbf{f}_{\mathrm{RF}, j}\mathbf{f}^H_{\mathrm{RF}, j} \Vert _{F}^{2} \right)^{2} =\lambda P^2_\mathrm{t},  
\end{aligned}
\end{equation}    
where $\mathbf{f}_{\mathrm{RF}, j}$ is the $j$th column of $\mathbf{F}_\mathrm{RF}$.
Thus, the surrogate function of the RBPS $\Psi(\mathbf{F}_\mathrm{RF})$ can be upper-bounded as
\begin{equation} \label{eqn:RF_7}
\mathbf{\Psi}(\mathbf{F}_\mathrm{RF})\leq \Re \lbrace \mathbf{d}^H\mathbf {b}^{t} \rbrace + c_{1}^{t} + \lambda P^2_\mathrm{t}.  
%\vspace{-1mm}
\end{equation}
Now, in support of our algorithm design, we rewrite the quantity $\Re \lbrace \mathbf{d}^H\mathbf {b}^{t} \rbrace$ in Eq. (\ref{eqn:RF_7}) equivalently as
\begin{equation} \label{eqn:RF_8}
\begin{aligned}
\Re \lbrace \mathbf{d}^H&\mathbf {b}^{t} \rbrace %=\sum _{j=1}^{M_\mathrm{t}}\Re \lbrace \text{vec}^{H}(\mathbf{f}_{\mathrm{RF}, j}\mathbf{f}^H_{\mathrm{RF}, j})\mathbf {b}^{t}\rbrace \\ 
=&\sum _{j=1}^{M_\mathrm{t}}\Re \lbrace \mathbf{f}^H_{\mathrm{RF}, j}\mathbf {B}^{t}\mathbf{f}_{\mathrm{RF}, j}\rbrace, 
\end{aligned} 
\end{equation}
where $\mathbf {B}^{t}\in \mathbb{C}^{N_\mathrm{t}\times N_\mathrm{t}}$ is a transformed equivalent of $\mathbf{b}^t$, which is defined as $\mathbf{b}^t = \text{vec}(\mathbf {B}^{t})$. According to (\ref{eqn:RF_3}) and (\ref{eqn:RF_5}), the matrix $\mathbf {B}^{t}$ can be further divided into two components as $\mathbf {B}^t\triangleq \mathbf {B}_1^t + \mathbf{B}_2^t$, where $\mathbf {B}_1^t$ and $\mathbf{B}_2^t$ are shown on the top of the next page.
\begin{figure*}[t]
\begin{align} \label{eqn:RF_9}
&\zeta \triangleq \sum _{l=1}^{L}G_{\mathrm{d}}^{2}(\theta_l),\\ 
&\mathbf {B}_{1}^{t} \!\triangleq\!  \frac{2}{L} \!\! \sum _{l_{1}=1}^{L}\frac{G_\mathrm{d}^{2}(\theta_{l_1})}{\zeta^2} \!\! \sum _{l_2=1}^{L}G_{\mathrm{d}}(\theta_{l_2})\mathbf{a}^H_{l_{2}} \!\mathbf{d}^t \sum _{l_3=1}^{L} G_\mathrm{d}(\theta_{l_3})\mathbf{A}_{l_{3}} +\frac{2}{L} \sum _{l_1=1}^{L}\text{vec}^{H}(\mathbf{A}_{l_1})\mathbf{d}^t\mathbf {A}_{l_1}, \label{eqn:RF_9_1}\\ 
&\mathbf {B}_{2}^{t}\triangleq  -\frac{4}{L}\Re \left\lbrace \! \sum _{l_{1}=1}^{L}\frac{G_\mathrm{d}(\theta_{l_1})}{\zeta} \mathbf{a}^H_{l_1}\mathbf{d}^t \times \sum_{l_2=1}^{L} G_{\mathrm{d}}(\theta_{l_2})\mathbf{A}_{l_2}\right\rbrace -2\lambda\mathbf{d}^t. \label{eqn:RF_9_2}
\end{align}
\normalsize
\hrulefill
\end{figure*}
Therefore, $(\ref{eqn:RF_8})$ can be rewritten as  
\begin{equation}
\begin{aligned}
&\Re \lbrace \mathbf{d}^H\mathbf {b}^{t} \rbrace
%&=\sum _{j=1}^{M_\mathrm{t}}\Re \lbrace \mathbf{f}^H_{\mathrm{RF}, j}\left(\mathbf {B}_{1}^{t}+\mathbf {B}_{2}^{t}\right)\mathbf{f}_{\mathrm{RF}, j}\rbrace,\\ 
=\sum _{j=1}^{M_\mathrm{t}}\Re \lbrace \mathbf{f}^H_{\mathrm{RF}, j}\mathbf {B}_{1}^{t}\mathbf{f}_{\mathrm{RF}, j}+\mathbf{f}^H_{\mathrm{RF}, j}\mathbf {B}_{2}^{t}\mathbf{f}_{\mathrm{RF}, j}\rbrace \label{eqn:RF_8_1}.
\end{aligned}
\end{equation}
Observe from (\ref{eqn:RF_9_1}) and (\ref{eqn:RF_9_2}) that the matrices $\mathbf {B}_{1}^{t}$ and $\mathbf{B}^t_2$ are positive- and negative-semidefinite matrices, respectively. Therefore, the first function $\mathbf{f}^H_{\mathrm{RF}, j}\mathbf {B}_{1}^{t}\mathbf{f}_{\mathrm{RF}, j}$ and the second function $\mathbf{f}^H_{\mathrm{RF}, j}\mathbf {B}_{2}^{t}\mathbf{f}_{\mathrm{RF}, j}$ of (\ref{eqn:RF_8_1}) are convex and concave functions, respectively. To further address the non-convex component of $\mathbf{f}^H_{\mathrm{RF}, j}\mathbf {B}_2^{t}\mathbf{f}_{\mathrm{RF}, j}$, we again adopt the MM approach to obtain the convex surrogate function for it. Specifically, by exploiting the Taylor series expansion, the concave function $\mathbf{f}^H_{\mathrm{RF}, j}\mathbf {B}_2^{t}\mathbf{f}_{\mathrm{RF}, j}$ can be upper bounded as   
\begin{equation}\label{eqn:RF_10}
\begin{aligned} 
\mathbf{f}^H_{\mathrm{RF}, j}\mathbf{B}_{2}^{t}\mathbf{f}_{\mathrm{RF}, j}\leq & (\mathbf{f}^t_{\mathrm{RF}, j})^{H}\mathbf {B}_{2}^{t}(\mathbf{f}^t_{\mathrm{RF}, j}) \\
&+2\Re \left\lbrace (\mathbf{f}^t_{\mathrm{RF}, j})^{H}\mathbf {B}_{2}^{t}\left(\mathbf{f}_{\mathrm{RF}, j}-(\mathbf{f}^t_{\mathrm{RF}, j})\right)\right\rbrace, 
\end{aligned}
\end{equation}
where the right hand side of (\ref{eqn:RF_10}) is the convex surrogate function of the concave function $\mathbf{f}^H_{\mathrm{RF}, j}\mathbf{B}_{2}^{t}\mathbf{f}_{\mathrm{RF}, j}$. Hence, by substituting (\ref{eqn:RF_8})-(\ref{eqn:RF_10}) into (\ref{eqn:RF_7}), a convex upper bound for the MSE function of the beampattern $\Psi(\mathbf{F}_\mathrm{RF})$ is given by
\begin{equation}\label{eqn:RF_11}
\Psi(\mathbf{F}_\mathrm{RF})\leq \sum _{j=1}^{M_\mathrm{t}} \Re \left\lbrace \mathbf{f}^H_{\mathrm{RF}, j}\mathbf{B}_{1}^t\mathbf{f}_{\mathrm{RF}, j} +2\mathbf{f}^H_{\mathrm{RF}, j}\mathbf{u}_{j}\right\rbrace +c_{2}^{t},
\end{equation}
where we have
\begin{align} 
\mathbf {u}_{j}&\triangleq ({\mathbf {B}_2^t})^H(\mathbf{f}^t_{\mathrm{RF}, j}) \label{eqn:RF_12},\\ 
c_{2}^{t}&\triangleq -\!\!\!\sum _{j=1}^{M_\mathrm{t}} \Re \lbrace ((\mathbf{f}^t_{\mathrm{RF}, j}))^H({\mathbf {B}_2^t})^H(\mathbf{f}^t_{\mathrm{RF}, j})\rbrace +c_{1}^{t} + \lambda M^2_\mathrm{t}\label{eqn:RF_12_1}. 
\end{align}
Upon substituting the convex surrogate function for the MSE of the RBPS  $\Psi(\mathbf{F}_\mathrm{RF})$ from Eq. (\ref{eqn:RF_11}) into (\ref{eqn:RF_1}), the optimization problem of updating $\mathbf{F}_\mathrm{RF}$ can be reformulated as
\begin{subequations}\label{eqn:RF_ADMM_1}
\begin{align}
&\mathop{\min}_{\mathbf{F}_\mathrm{RF}}\|\mathbf{F}^\mathrm{opt}-\mathbf{F}_\mathrm{RF}\widetilde{\mathbf{F}}^{\mathrm{C}, 1}_\mathrm{BB}\|^2_F \\
&\text {s.~t.} ~
\sum_{j=1}^{M_\mathrm{t}} \Re \left\lbrace \mathbf{f}^H_{\mathrm{RF}, j}\mathbf{B}_{1}^t\mathbf{f}_{\mathrm{RF}, j} +2\mathbf{f}^H_{\mathrm{RF}, j}\mathbf{u}_{j}\right\rbrace +c_{2}^{t} \leq \epsilon \\
&\qquad~\left\vert\mathbf{F}_\mathrm{RF}(i,j)\right\vert = \frac{1}{\sqrt{N_\mathrm{t}}}, \forall i, j. \label{constr:RF_4}
\end{align}
\end{subequations}

The above problem is still non-convex due to the constant modulus constraint imposed on each element of $\mathbf{F}_\mathrm{RF}$.
This problem can be solved by applying the SDR technique to relax the non-convex constant modulus constraint. Usually, the SDR technique requires an additional randomization step, since it may provide an infeasible solution due to the rank-one constraint. Moreover, the SDR technique may fail to find a feasible solution, when the number of CUs and targets equals the number of RF chains. To effectively handle this problem, we adopt the popular RCG based method to solve (\ref{eqn:RF_ADMM_1}), which is discussed next.

Let us define the feasible set $\mathcal{F}$ for (\ref{eqn:RF_ADMM_1}) on the complex circle manifold as
\begin{equation} 
\begin{aligned}
\mathcal {F} = \begin{cases} &\mathbf{F}\in \mathbb{C}^{N_\mathrm{t}\times M_\mathrm{t}} \bigg| 
\sum_{j=1}^{M_\mathrm{t}} \Re \left\lbrace \mathbf{f}^H_{\mathrm{RF}, j}\mathbf{B}_{1}^t\mathbf{f}_{\mathrm{RF}, j} \right.\\
&\left.+2\mathbf{f}^H_{\mathrm{RF}, j}\mathbf{u}_{j}\right\rbrace 
+c_{2}^{t} \leq \epsilon, \\
&\vert\mathbf{F}(i,j)\vert=\frac{1}{\sqrt{N_\mathrm{t}}}, \forall i,j. 
\end{cases}
\end{aligned}
\end{equation}
Thus, the optimization problem (\ref{eqn:RF_ADMM_1}) can be recast as
\begin{equation}
\begin{aligned}
&\min_{\mathbf{F}_\mathrm{RF}}f(\mathbf{F}_\mathrm{RF})= \|\mathbf{F}^\mathrm{opt}-\mathbf{F}_\mathrm{RF}\widetilde{\mathbf{F}}^{\mathrm{C}, 1}_\mathrm{BB}\|^2_F ~~{\text {s.~t. }} ~ \mathbf{F}_\mathrm{RF}(i,j) \in \mathcal{F}.
\end{aligned}
\end{equation}
Furthermore, the Euclidean gradient of $f(\mathbf{F}_\mathrm{RF})$ is given by
\begin{equation}\label{RCG_3}
\begin{aligned}
\nabla f(\mathbf{F}_\mathrm{RF})=-2\left(\mathbf{F}^\mathrm{opt}\left(\widetilde{\mathbf{F}}^{\mathrm{C}, 1}_\mathrm{BB}\right)^H + \mathbf{F}_\mathrm{RF}\widetilde{\mathbf{F}}^{\mathrm{C}, 1}_\mathrm{BB}\left(\widetilde{\mathbf{F}}^{\mathrm{C}, 1}_\mathrm{BB}\right)^H\right).
\end{aligned}
\end{equation}
However, the RCG algorithm employs the Riemannian gradient to evaluate the descent direction, which is defined as the orthogonal projection of $\nabla f(\mathbf{F}_\mathrm{RF})$ onto the tangent space $T_{\mathbf{F}^i_\mathrm{RF}}\mathcal{F}$ of the manifold $\mathcal{F}$ at the associate point $\mathbf{F}^i_\mathrm{RF}$. This is given as
\begin{equation}\label{RCG_2}
T_{\mathbf{F}^i_\mathrm{RF}} \mathcal {F}=\left \{{\mathbf{F}_\mathrm{RF} \in \mathbb {C}^{N_\mathrm{t} \times M_\mathrm{t}}: \Re\left \{{\mathbf{F}_\mathrm{RF} \odot (\mathbf{F}^i_\mathrm{RF})^{*}}\right \}=\mathbf {0}_{N_\mathrm{t} \times M_\mathrm{t}}}\right \}.
\end{equation}
Subsequently, the Riemannian gradient at the point $\mathbf{F}^i_\mathrm{RF}$ is obtained as
\begin{equation} \label{RCG_4}
\begin{aligned}
\mathrm {grad}~ f(\mathbf{F}^i_\mathrm{RF})=\nabla f(\mathbf{F}^i_\mathrm{RF}) -\Re\left \{{\nabla f(\mathbf{F}^i_\mathrm{RF}) \odot (\mathbf{F}^i_\mathrm{RF})^{*}}\right \} \odot \mathbf{F}^i_\mathrm{RF}.
\end{aligned}
\end{equation}
Similar to the conjugate gradient method of the Euclidean space, the update rule of the search direction in the manifold space is given by 
\begin{equation} \label{RCG_5}
\begin{aligned}
\boldsymbol {\Xi }^{i+1}=-\mathrm {grad}~ f(\mathbf{F}^{i+1}_\mathrm{RF})+\nu \mathcal {T}_{\mathbf{F}^i_\mathrm{RF} \rightarrow \mathbf{F}^{i+1}_\mathrm{RF}}\left ({\boldsymbol {\Xi }^{i}}\right),
\end{aligned}
\end{equation}
where $\boldsymbol {\Xi }^{i}$ denotes the search direction at $\mathbf{F}^i_\mathrm{RF}$, $\nu$ is the update parameter chosen as the Polak-Ribiere parameter \cite{application_4,application_5}, and $\mathcal {T}_{\mathbf{F}^i_\mathrm{RF} \rightarrow \mathbf{F}^{i+1}_\mathrm{RF}}\left ({\boldsymbol{\Xi}^{i}}\right)$ represents the transport operation, which is required because both $\boldsymbol {\Xi }^{i+1}$ and $\boldsymbol {\Xi }^i$ are in different tangent spaces and operations such as the sum in (\ref{RCG_5}) cannot be carried out directly. Therefore, the transport operation $\mathcal {T}_{\mathbf{F}^i_\mathrm{RF} \rightarrow \mathbf{F}^{i+1}_\mathrm{RF}}\left ({\boldsymbol {\Xi}^{i}}\right)$ is required to map the tangent vector at the previous search direction to its original tangent space at the current point $\mathbf{F}^{i+1}_\mathrm{RF}$, which is formulated as
\begin{equation}\label{RCG_6}
\begin{aligned} 
\mathcal {T}_{\mathbf{F}^{i}_\mathrm{RF} \rightarrow \mathbf{F}^{i+1}_\mathrm{RF}}\left ({\boldsymbol {\Xi }^{i}}\right): & T_{\mathbf{F}^{i}_\mathrm{RF}} \mathcal {F} \mapsto T_{\mathbf{F}^{i+1}_\mathrm{RF}} \mathcal {F}: \\
& \,\,\,\, \boldsymbol {\Xi}^{i} \mapsto \boldsymbol {\Xi}^{i}-\Re\left \{{\boldsymbol {\Xi }^{t} \odot (\mathbf{F}^{i+1}_\mathrm{RF})^{*}}\right \} \odot \mathbf{F}^{i+1}_\mathrm{RF}.
\end{aligned}
\end{equation}
After determining the search direction $\boldsymbol {\Xi }^{i+1}$, the retraction operation $\mathrm {Retr}_{\mathbf{F}^{i}_\mathrm{RF}}(\kappa \boldsymbol {\Xi }^i)$ is performed to find the destination on the manifold. Specifically,  $\mathrm {Retr}_{\mathbf{F}^{i}_\mathrm{RF}}(\kappa \boldsymbol {\Xi}^i)$ maps the point on the tangent space $T_{\mathbf{F}^{i}_\mathrm{RF}}\mathcal{F}$ to the manifold $\mathcal{F}$, which is given by
\begin{equation}\label{RCG_7}
\begin{aligned} 
\mathrm {Retr}_{\mathbf{F}^{i}_\mathrm{RF}}&(\kappa \boldsymbol {\Xi}^i):\\
&T_{\mathbf{F}^{i}_\mathrm{RF}} \mathcal {F}\mapsto&\mathcal {F}:
\kappa \boldsymbol {\Xi }^{i}\mapsto \mathrm{Pj}_{\mathcal{F}}\Big(\left ({\mathbf{F}^{i}_\mathrm{RF}+\kappa \boldsymbol {\Xi}^{i}}\right)\Big),
\end{aligned}
\end{equation}
where $\kappa$ is the Armijo backtracking line search step size \cite{application_6} and $\mathrm{Pj}$ denotes the projection operation. The key steps of the RMCG algorithm proposed above to solve the problem (\ref{eqn:RF_1}) are summarized in Algorithm 1.
\begin{algorithm}[t]
\caption{RMCG algorithm for solving (\ref{eqn:RF_1}) }
 \textbf{Input:} Optimal TPC $\mathbf{F}^\mathrm{opt}\in \mathbb{C}^{N_\mathrm{t}\times M}$, baseband TPC $\widetilde{\mathbf{F}}^{\mathrm{C}, 1}_\mathrm{BB} \in \mathbb{C}^{M_\mathrm{t}\times M}$, stopping parameters $\varrho$ and $\varsigma$\\
 \textbf{Output:} RF TPC $\mathbf{F}_\mathrm{RF}$
\begin{algorithmic}[1]\label{alg:MM_RCG_algo}
\State \textbf{initialize:} $\mathbf{F}^0_\mathrm{RF}\in \mathcal{F}$, outer iteration $t=0$
\While{$\left(\|\mathbf{F}^\mathrm{opt}-\mathbf{F}_\mathrm{RF}\widetilde{\mathbf{F}}^{\mathrm{C}, 1}_\mathrm{BB}\|^2_F \leq \varsigma \right)$}
   \State Obtain $\beta$ using (\ref{eqn:beta})
   \State Compute $\mathbf {B}_{1}^{t}$ and $\mathbf {B}_{2}^{t}$ using (\ref{eqn:RF_9_1}) and (\ref{eqn:RF_9_2})
   \State Evaluate $\mathbf {u}_{j}$ and $c_{2}^{t}$ using (\ref{eqn:RF_12}) and (\ref{eqn:RF_12_1})
   \State Obtain $\Psi(\mathbf{F}^{t+1}_\mathrm{RF})=\sum _{j=1}^{M_\mathrm{t}} \Re \left\lbrace \mathbf{f}^H_{\mathrm{RF}, j}\mathbf{B}_{1}^t\mathbf{f}_{\mathrm{RF}, j} +2\mathbf{f}^H_{\mathrm{RF}, j}\mathbf{u}_{j}\right\rbrace +c_{2}^{t}$
   \State \textbf{initialize:} $\mathbf{F}^0_\mathrm{RF}=\mathbf{F}^t_\mathrm{RF}$
    \State Calculate $\boldsymbol{\Xi}^0 = -\mathrm {grad}~ f(\mathbf{F}^0_\mathrm{RF})$ according to (\ref{RCG_4}) and set inner iteration $i=0$;
    \While{$\left(\left \vert \left \vert\mathrm {grad}~ f(\mathbf{F}^i_\mathrm{RF}) \right\vert\right\vert_2 \leq \varrho \right)$}
        \State Choose the Armijo backtracking line search step size $\kappa$
        \State Obtain the next point $\mathbf{F}^{i+1}_\mathrm{RF}$ using retraction in (\ref{RCG_7})
        \State Determine Riemannian gradient $\mathrm {grad}~ f(\mathbf{F}^{i+1}_\mathrm{RF})$ according to (\ref{RCG_4})
        \State Obtain the transport $\mathcal{T}_{\mathbf{F}_\mathrm{RF}^i \rightarrow \mathbf{F}^{i+1}_\mathrm{RF}}\left(\boldsymbol{\Xi}^i\right)$ according to (\ref{RCG_6})
        \State Choose the Polak-Ribiere parameter $\nu$
        \State Calculate the conjugate direction $\boldsymbol{\Xi}^{i+1}$ according to (\ref{RCG_5})
        \State $i\leftarrow i+1$
    \EndWhile
    \State \textbf{end}
    \State \textbf{upate:} $\mathbf{F}^{t+1}_\mathrm{RF}=\mathbf{F}^{i}_\mathrm{RF}$ and $\widetilde{\mathbf{F}}^{\mathrm{C}, 1}_\mathrm{BB}$ using (\ref{eqn:F_BB1_tilde})
    \State \textbf{upate:} $\beta$ and $\widetilde{\mathbf{F}}^{\mathrm{C}, 1}_\mathrm{BB}$ using (\ref{eqn:F_BB1_tilde}) and (\ref{eqn:beta})
    \State $t\leftarrow t+1$
    \EndWhile
    \State \textbf{end}
    \State \textbf{return:} $\mathbf{F}_\mathrm{RF}=\mathbf{F}^{t+1}_\mathrm{RF}$ 
\end{algorithmic}
\end{algorithm}
%\vspace{-2mm}
\subsubsection{Digital baseband beamformer design}\label{Digital baseband beamformer design}
In this subsection, we aim for designing the baseband TPC at the ISAC BS, which maximizes the sum-SE of the CUs as well as mitigates the MUI. As discussed in the RF TPC design procedure, we follow an approach to design the RBPS of the ISAC BS that depends only on the RF TPC $\mathbf{F}_\mathrm{RF}$, which constrains the baseband TPC to be a unitary matrix. 
To this end, let us now first determine the sub-matrices of the unconstrained baseband TPC, which is given as $\mathbf{F}_\mathrm{BB} = \left[\mathbf{F}^\mathrm{C}_\mathrm{BB} \hspace{2mm}\mathbf{F}^\mathrm{R}_\mathrm{BB}\right]$, where $\mathbf{F}^\mathrm{C}_\mathrm{BB}=\widetilde{\mathbf{F}}^{\mathrm{C}, 1}_\mathrm{BB}\widetilde{\mathbf{F}}^{\mathrm{C}, 2}_\mathrm{BB}$. For a fixed RF TPC, one can obtain $\widetilde{\mathbf{F}}^{\mathrm{C}, 1}_\mathrm{BB}$ in the first-stage of TPC design from equation (\ref{eqn:RF_1}) using the least squares solution, which is given by 
\begin{equation}\label{eqn:F_BB1_tilde}
\begin{aligned}
\widetilde{\mathbf{F}}^{\mathrm{C}, 1}_\mathrm{BB}=\left(\mathbf{F}^H_{\rm RF}\mathbf{F}_{\rm RF}\right)^{-1}\mathbf{F}^H_{\rm RF}\mathbf{F}^\mathrm{opt}.
\end{aligned}
\end{equation}
Additionally, to mitigate the MUI from the CUs, we design $\widetilde{\mathbf{F}}^{\mathrm{C}, 2}_\mathrm{BB}$ in the second-stage of the baseband TPC using the ZF method. According to the ZF technique, $\widetilde{\mathbf{F}}^{\mathrm{C}, 2}_\mathrm{BB}$ is given by
\begin{equation}\label{eqn:F_BB2_tilde}
\begin{aligned}
\widetilde{\mathbf{F}}^{\mathrm{C}, 2}_\mathrm{BB}=\widetilde{\mathbf{H}}^H_m\left(\widetilde{\mathbf{H}}_m\widetilde{\mathbf{H}}^H_m\right)^{-1},
\end{aligned}
\end{equation}
where $\widetilde{\mathbf{H}}_m=\mathbf{H}_m \mathbf{F}_\mathrm{RF}\widetilde{\mathbf{F}}^{\mathrm{C}, 1}_\mathrm{BB}\in \mathbb{C}^{N_\mathrm{r}\times M}$. Furthermore, as seen from $(\ref{eqn:rx signal_2})$, there is some additional interference due to the RTs, which is owing to the term $ \mathbf{F}_{\rm RF}\mathbf{F}^\mathrm{R}_\mathrm{BB}$. To mitigate this interference due to the RTs, we adopt the null-space projection (NSP) technique such that $\mathbf{H}_m\mathbf{F}_\mathrm{RF}\mathbf{F}^\mathrm{R}_\mathrm{BB}\approx \mathbf{0}$. According to the NSP method, we first perform the SVD of $\overline{\mathbf{H}}_m = \mathbf{H}_m\mathbf{F}_\mathrm{RF} \in \mathbb{C}^{N_\mathrm{r}\times M_\mathrm{t}}$ as $\overline{\mathbf{H}}_m = \overline{\mathbf{U}}_m\overline{\mathbf{\Sigma}}_m\overline{\mathbf{V}}^H_m$, where $\overline{\mathbf{U}}_m\in\mathbb{C}^{N_{\rm r}\times N_{\rm r}}, \overline{\mathbf{V}}_m\in\mathbb{C}^{M_{\rm t}\times M_{\rm t}}$ are the left and right singular matrices, while $\overline{\mathbf{\Sigma}}_m\in\mathbb{C}^{N_{\rm r}\times M_{\rm t}}$ is the diagonal matrix containing the singular values.
Consequently, $\mathbf{F}^\mathrm{R}_\mathrm{BB}$ is given by     
\begin{equation}\label{eqn:F_BB2_1}
\begin{aligned}
\mathbf{F}^\mathrm{R}_\mathrm{BB}=\overline{\mathbf{V}}_{m}(:,M_\mathrm{t}-L+1:M_\mathrm{t}).
\end{aligned}
\end{equation}
After obtaining the unconstrained baseband TPC $\mathbf{F}_\mathrm{BB}$, the corresponding constrained baseband TPC design problem is given by 
\begin{subequations}\label{eqn:F_BB2_2}
\begin{align}
%\vspace{-3mm}
\mathcal{P}_2: \hspace{10mm} &\min \limits_{\widehat{\mathbf{F}}_\mathrm{BB}} \quad  f\left(\widehat{\mathbf{F}}_\mathrm{BB}\right) =\left \|\mathbf{F}_\mathrm{BB}-\widehat{\mathbf{F}}_\mathrm{BB}\right \|_{F}^{2} \\
&\text{s.t.} \quad~\widehat{\mathbf{F}}_\mathrm{BB}\widehat{\mathbf{F}}^H_\mathrm{BB} = \frac{P_T}{K}\mathbf {I}_K,
\end{align}
%\vspace{-3mm}
\end{subequations}
where the constraint in the above optimization problem maintains the total transmit power at the ISAC BS. Note that the above optimization problem is still non-convex due to the $K$ quadratic equality constraints. While it can potentially also be solved via the SVD-based orthogonal Procrustes problem (OPP) \cite{ISAC_mmWave_35}, for ISAC BS systems we adopt the RCG framework for better efficiency, accuracy, and scalability in hybrid TPC design \cite{new_102}.
%While it can be solved via the SVD-based orthogonal Procrustes problem (OPP) \cite{ISAC_mmWave_35}, we adopt the RCG framework for better efficiency, accuracy, and scalability in hybrid TPC design for ISAC BS systems \cite{new_102}.} 
To this end, the Euclidean gradient of the function $f\left(\widehat{\mathbf{F}}_\mathrm{BB}\right)$ is given by $\nabla f\left(\widehat{\mathbf{F}}_\mathrm{BB}\right)=2\left(\mathbf{F}_\mathrm{BB}-\widehat{\mathbf{F}}_\mathrm{BB}\right)$. Therefore, the problem (\ref{eqn:F_BB2_2}) can be effectively solved again by the RCG algorithm upon changing the retraction operation as follows 
\begin{equation}\label{RCG_7_1}
\begin{aligned} 
&\mathrm {Retr}_{\mathbf{F}^i_\mathrm{BB}}(\kappa \boldsymbol {\Xi}): \\
&\sqrt{\frac{P_T}{K}}\left(\left ({\mathbf{F}_\mathrm{BB}+\kappa \boldsymbol {\Xi}}\right)\left ({\mathbf{F}_\mathrm{BB}+\kappa \boldsymbol {\Xi}}\right)^H\right)^{-1/2}\left ({\mathbf{F}_\mathrm{BB}+\kappa \boldsymbol {\Xi}}\right).
\end{aligned}
\end{equation}
\subsection{Receive combiner design}\label{blind MMSE}
After obtaining the hybrid TPC, we have to determine the RC at each CU. Therefore, the pertinent optimization problem is given by
%\vspace{-2mm}
\begin{subequations}\label{eqn:RC_1}
\begin{align}
&\mathop{\max}_{\bigl\{\mathbf{w}_{m}\bigr\}_{m=1}^M}\sum_{m=1}^{M}R_m\left( \mathbf{w}_m\right)~~ \text {s.~t.} \quad \text{(\ref{constr:comb_1}).}
\end{align}
%\vspace{-5mm}
\end{subequations}
To address this problem, we focus our attention on the design of a linear RC for the system, which is more practical. However, the authors of \cite{ISAC_mmWave_14,ISAC_mmWave_13,ISAC_mmWave_12} proposed a linear RC design that relies on both the CSI and on the complete knowledge of the RF and baseband TPCs  $\mathbf{F}_\mathrm{RF},\mathbf{F}_\mathrm{BB}$. These techniques are inefficient due to the requirement of a large feedback overhead to convey $\mathbf{F}_\mathrm{RF},\mathbf{F}_\mathrm{BB}$ from the ISAC BS to all the CUs.
To circumvent this impediment, we propose an efficient low-complexity linear MM-based blind combiner (LMBC) design for each CU, which does not rely on the knowledge of the RF and baseband TPC. 

To this end, each CU designs the MMSE combiner considering the optimal fully digital TPC being used at the ISAC BS. Upon setting $\mathbf{F}_\mathrm{RF}\mathbf{F}_\mathrm{BB}=\mathbf{F}^\mathrm{opt}$ and employing $\mathbf{F}^\mathrm{R}_\mathrm{BB}$ obtained using the NSP technique described in Section \ref{Digital baseband beamformer design}, the signal received at the $m$th CU and formulated in (\ref{eqn:rx signal_2}) is given by
\begin{subequations}\label{eqn:RC_2}
\begin{align}
\Bar{\mathbf {y}}_{m}&=\mathbf{H}_m\Bar{\mathbf{F}}^\mathrm{opt}\mathbf{x}_1 + \mathbf{n}_{m},\\
&=\mathbf{H}_{m} \Bar{\mathbf{f}}^\mathrm{opt}_m s_{m} +\sum_{n=1, n \neq m}^M  
  \mathbf{H}_{m} \Bar{\mathbf{f}}^\mathrm{opt}_n s_n + \mathbf{n}_{m}, 
\end{align}
\end{subequations}
where we have $\Bar{\mathbf{F}}^\mathrm{opt} = [\Bar{\mathbf{f}}^\mathrm{opt}_1, \hdots, \Bar{\mathbf{f}}^\mathrm{opt}_{M}]$. The first and second terms in the above equation are the desired and MUI signals. Hence, we aim to design $\Bar{\mathbf{F}}^\mathrm{opt}$, which can mitigate the MUI and enhance the desired signal power. To design this TPC we once again consider the SVD of $\mathbf{H}_m$, which is given by
\begin{equation}\label{eqn:RC_3}
\begin{aligned}
\Bar{\mathbf{f}}^\mathrm{opt}_m =\mathbf{V}_{m}(:,1), \Bar{\mathbf{f}}^\mathrm{opt}_{i,i\neq m} =\mathbf{V}_{m}(:,N_\mathrm{t}-M+1:N_\mathrm{t}),
\end{aligned}
\end{equation}
where $\mathbf{V}_{m}(:,1)\in \mathbb{C}^{N_\mathrm{t}\times 1}$ is the singular vector corresponding to the highest singular value of $\mathbf{H}_m$.
Based on the above $\Bar{\mathbf{F}}^\mathrm{opt}$ design, the received signal $(\ref{eqn:RC_2})$ can be rewritten as
\begin{equation}\label{eqn:RC_4}
\begin{aligned}
\Bar{\mathbf {y}}_{m}\approx&\mathbf{U}_m(:,1)\mathbf{\Sigma}_{m}(1,1)s_m + \mathbf{n}_{m}.
\end{aligned}
\end{equation}
\begin{algorithm}[t]
\caption{LMBC algorithm for receive combiner design}
\label{alg:algo_2}
\begin{algorithmic}[1]
\Require $\mathbf{H}_m$,  stopping parameter $\vartheta$  
\For{$m=1:M$ }
    \State Initialize $\Phi_m$
    \State Compute SVD of $\mathbf{H}_m$ as $\mathbf{H}_m=\mathbf{U}_{m}\mathbf{\Sigma}_{m}\mathbf{V}^H_{m}$.
    \State Obtain $\Bar{\mathbf{f}}^\mathrm{opt}_m =\mathbf{V}_{m}(:,1), \Bar{\mathbf{f}}^\mathrm{opt}_{i,i\neq m} =\mathbf{V}_{m}(:,N_\mathrm{t}-M+1:N_\mathrm{t})$
    \State Construct $\Bar{\mathbf{F}}^\mathrm{opt} = [\Bar{\mathbf{f}}^\mathrm{opt}_1, \hdots, \Bar{\mathbf{f}}^\mathrm{opt}_{M}]$
    \State Obtain $\Bar{\mathbf {y}}_{m}$ using (\ref{eqn:RC_2})
           \While{$g\left(\Phi_m\right) \leq \vartheta $}
           \State Obtain $g\left(\Phi_m^t\right)$ using (\ref{eqn:RC_1_4})
           \State Compute the step-size $\varkappa$ using the Armijo rule 
           \State \textbf{upate:} $\Phi_m^{t+1} = \Phi_m^{t} - \varkappa \nabla g(\Phi_m^{t})$
           \State $t\leftarrow t+1$
           \EndWhile
           \State \textbf{end}
    \EndFor
    \State $\mathbf{w}_m = \frac{1}{\sqrt{N}_r}e^{j\Phi_m}$
    \State \textbf{end}
\State $\mathbf{return}$ \hspace{3pt} $\mathbf{w}_m, \forall m$ 
\end{algorithmic}
\end{algorithm}
The linear RC design, which minimizes the MSE of the transmitted and received signal is given by  
\begin{equation}\label{eqn:RC_1_2}
\begin{aligned}
&\min \limits_{\mathbf{w}_m} \quad  f\left(\mathbf{w}_m\right)= \sum_{m=1}^M\mathbb{E}\left[\left\vert s_m-
\mathbf{w}_m^H \Bar{\mathbf{y}}_m\right\vert_2^2\right]~~\text {s.~t.} ~ \text{(\ref{constr:comb_1}).}
\end{aligned}
\end{equation}
The above optimization problem is non-convex due to the non-convex nature of the objective function and constant modulus constraint imposed on the elements of $\mathbf{w}_m$. To solve this problem, we first rewrite the above problem as an unconstrained non-convex optimization problem
\begin{equation} \label{eqn:RC_1_3}
\mathcal{P}_3: \hspace{10mm}\min \limits_{ \Phi_m } \quad g(\Phi_m)=\sum \limits_{m=1}^{M} \mathbb{E}\left[\left |s_m-{ (e^{j \Phi_m})^{H}\Bar{\mathbf{y}}_{m} }\right |^{2}\right],
\end{equation}
where $ \Phi_m=\angle{\mathbf{w}_m}$. However, due to the non-convex objective function, the above optimization problem is still challenging to solve. Therefore, we again exploit the MM technique, which successively approximates the non-convex objective function by a convex one using its surrogate function. Let us upper-bound $g(\Phi_m)$ using its surrogate function at its current local point $g(\Phi^t)$ in the $t$th iteration using the second-order Taylor expansion as
\begin{equation} \label{eqn:RC_1_4}
g(\Phi_m)\leq g(\Phi_m^t)+\nabla g(\Phi_m^t)^{T} (\Phi_m -\Phi_m^t )+\frac {1}{2\varkappa }\| \Phi_m - \Phi_m^t\|^{2},
\end{equation}
where $\varkappa$ is chosen to satisfy $g(\Phi_m)\leq g(\Phi_m^t)$ within a bounded feasible set. In practice $\varkappa$ is computed using the Armijo rule, which is given as
\begin{equation} 
g(\Phi_m^t)- g(\Phi_m^{t+1}) \geq \iota \varkappa \|\nabla g(\Phi_m^t) \|^{2},
\end{equation}
where  $o<\iota<0.5$. Subsequently, we update $\Phi_m$ as
\begin{equation} \label{eqn:RC_1_5}
\Phi_m^{t+1} = \Phi_m^{t} - \varkappa \nabla g(\Phi_m^{t}).
\end{equation}
The key steps of the LMBC algorithm discussed above to design the RC are summarized in Algorithm 2. Furthermore, the proposed sequential algorithm for designing the hybrid TPC and RC is discussed in Algorithm 3.
\begin{algorithm}[t]
\caption{Proposed sequential algorithm to design the hybrid transceiver for mmWave MIMO ISAC systems}
\label{alg:algo_3}
\begin{algorithmic}[1]
\Require $\mathbf{H}_m$, stopping parameters $\varrho$, $\varsigma$ and $\vartheta$     
    \State Optimize $\mathbf{F}_\mathrm{RF}$ by RMCG Algorithm 1.
    \State Compute $\widetilde{\mathbf{F}}^{\mathrm{C}, 1}_\mathrm{BB}, \widetilde{\mathbf{F}}^{\mathrm{C}, 2}_\mathrm{BB}$ and $\mathbf{F}^\mathrm{R}_\mathrm{BB}$ using equations (\ref{eqn:F_BB1_tilde}), (\ref{eqn:F_BB2_tilde}) and (\ref{eqn:F_BB2_1}), respectively
    \State Obtain the unconstrained baseband TPC as $\mathbf{F}_\mathrm{BB} = \left[\widetilde{\mathbf{F}}^{\mathrm{C}, 1}_\mathrm{BB}\widetilde{\mathbf{F}}^{\mathrm{C}, 2}_\mathrm{BB} \hspace{2mm}\mathbf{F}^\mathrm{R}_\mathrm{BB}\right]$
    \State Optimize the baseband TPC $\widehat{\mathbf{F}}_\mathrm{BB}$ by solving (\ref{eqn:F_BB2_2}) using the RCG algorithm
    \State Optimize the RC using the LMBC Algorithm 2
\State $\mathbf{return} \hspace{3pt} \mathbf{F}_\mathrm{RF},\mathbf{F}_\mathrm{BB}$ and $\mathbf{w}_m, \forall m$ 
\end{algorithmic}
\end{algorithm}
We now discuss the overall computational complexity of the proposed sequential algorithm. While designing the RF TPC via the iterative RMCG algorithm, the complexities for obtaining $\mathbf{B}_\mathrm{1}$ and $\mathbf{B}_\mathrm{2}$ are dominant in the outer loop, and both are given by $\mathcal{O}(L^2N^3_\mathrm{t}M_\mathrm{t})$. Furthermore, the computational cost of the inner loop in the RMCG algorithm is dominated by the Euclidean gradient, which is given by  $\mathcal{O}({N_\mathrm{t}M_\mathrm{t}M})$. Therefore, the overall complexity of the RMCG algorithm is given by $\mathcal{O}\bigg(I_\mathrm{out}\big(2LN^3_\mathrm{t}M_\mathrm{t}+I_\mathrm{in}\left(N_\mathrm{t}M_\mathrm{t}M\right)\big)\bigg)$, where $I_\mathrm{in}$ and $I_\mathrm{out}$ are the number of iterations requires in the inner and outer layer for the convergence. Furthermore, the costs for obtaining $\widetilde{\mathbf{F}}^{\mathrm{C}, 1}_\mathrm{BB}$ and $\widetilde{\mathbf{F}}^{\mathrm{C}, 2}_\mathrm{BB}$ via least squares and ZF techniques are given by $\mathcal{O}(2M^3_\mathrm{t}MN_\mathrm{t})$ $\mathcal{O}(2MN^3_\mathrm{r})$, respectively. On the other hand, the complexity for constructing RC $\mathbf{w}_m, \forall m,$ via the LMBC algorithm is $\mathcal{O}(I_\mathrm{o}MN_\mathrm{r})$, where $I_\mathrm{o}$ is the iteration required in the Armijo search. Therefore, the complexity of the overall sequential method is given by $\mathcal{O}\bigg(I_\mathrm{out}\big(2LN^3_\mathrm{t}M_\mathrm{t}+I_\mathrm{in}\left(N_\mathrm{t}M_\mathrm{t}M\right)\big) + 2M^3_\mathrm{t}MN_\mathrm{t} + 2MN^3_\mathrm{r} + I_\mathrm{o}MN_\mathrm{r}\bigg)$.

\section{Geometric mean of spectral efficiency maximization}
Although the hybrid transceiver design in the previous sections maximizes the sum-SE of the system, it also creates the problem of rate unfairness among the CUs. To handle the problem of rate unfairness, we consider the GM-SE metric for communication, which maximizes the GM of the CU's rate. Consequently, the optimization problem that maximizes the GM-SE of the system and meets the MSE tolerance of the RBPS and transmits power constraints is formulated as
\begin{equation}\label{eqn:GM_1}
\begin{aligned} 
&\max _{\bigl\{\mathbf{w}_{m}\bigr\}_{m=1}^M, \mathbf{F}_\mathrm{RF}, \mathbf{F}_\mathrm{BB}} \left ({\prod_{m=1}^{M} R_{m}\left(\mathbf{w}_{m}, \mathbf{F}_\mathrm{RF}, \mathbf{F}_\mathrm{BB}\right)}\right)^{1/M} \\ 
 & \text {s. t.}\quad \text{(\ref{constr:RBPS_1}), (\ref{constr:RF_1}), (\ref{constr:comb_1}) and (\ref{constr:TP_1})}.
\end{aligned} 
\end{equation}
Considering $\mathbf{w} \triangleq \{\mathbf{w}_m, \forall m\}$, the above problem (\ref{eqn:GM_1}) can be equivalently written as
\begin{equation}\label{eqn:GM_2}
\begin{aligned}
\mathcal{P}_4:\hspace{-1mm} &\hspace{-1mm}\min_{\mathbf{w}, \mathbf{F}_\mathrm{RF}, \mathbf{F}_\mathrm{BB}} f\big(R_{1}\left(\mathbf{w}, \mathbf{F}_\mathrm{RF}, \mathbf{F}_\mathrm{BB}\right),{\hdots},R_{M}\left(\mathbf{w}, \mathbf{F}_\mathrm{RF}, \mathbf{F}_\mathrm{BB}\right)\big) \\
&\triangleq \frac {1}{\left ({\prod_{m=1}^{M} R_{m}(\mathbf{w}, \mathbf {F}_\mathrm{RF}, \mathbf{F}_\mathrm{BB})}\right)^{1/M}} \\
& \text {s.~ t.}\quad \text{(\ref{constr:RBPS_1}), (\ref{constr:RF_1}), (\ref{constr:comb_1}) and (\ref{constr:TP_1})}.
\end{aligned} 
\end{equation}
Since the objective function of (\ref{eqn:GM_2}) %$f\left(R_1\left(\mathbf{w}, \mathbf{F}_\mathrm{RF}, \mathbf{F}_\mathrm{BB}\right), \hdots, R_M\left(\mathbf{w}, \mathbf{F}_\mathrm{RF}, \mathbf{F}_\mathrm{BB}\right)\right)$ 
is the composition of the convex function $f(R_1,\hdots,R_M)=\frac{1}{\left(\prod_{m=1}^{M}R_m \right)^{1/M}}$ and the non-convex 
functions $R_m\left(\mathbf{w}, \mathbf{F}_\mathrm{RF}, \mathbf{F}_\mathrm{BB}\right), \forall m$, it is highly intractable. Moreover, the non-convex constraints (\ref{constr:RBPS_1}), (\ref{constr:RF_1}) and (\ref{constr:comb_1}) exacerbate the challenge to solve the problem (\ref{eqn:GM_2}). 
To solve this problem, we first adopt the steepest descent approach to handle the objective function \cite{GM_1}.  
Let us assume that $(\mathbf{w}^{(\kappa)}, \mathbf{F}^{(\kappa)}_\mathrm{RF}, \mathbf{F}^{(\kappa)}_\mathrm{BB})$ is a feasible point for (\ref{eqn:GM_2}), which is obtained from the $(\kappa-1)$th iteration. Therefore, we write the non-linear function $f\left(R_1\left(\mathbf{w}, \mathbf{F}_\mathrm{RF}, \mathbf{F}_\mathrm{BB}\right), {\dots }, R_{M}\left(\mathbf{w}, \mathbf{F}_\mathrm{RF}, \mathbf{F}_\mathrm{BB}\right)\right)$ as a linear function via following (\ref{eqn:lin}). 
\begin{figure*}[t]
\begin{align}\label{eqn:lin}
&f\left(R_{1}\left(\mathbf{w}, \mathbf{F}_\mathrm{RF}, \mathbf{F}_\mathrm{BB}\right), {\dots }, R_{M}\left(\mathbf{w}, \mathbf{F}_\mathrm{RF}, \mathbf{F}_\mathrm{BB}\right)\right) \nonumber\\
&=2f\!\left(R_1\left(\mathbf{w}^{(\kappa)}, \mathbf{F}^{(\kappa)}_\mathrm{RF}, \mathbf{F}^{(\kappa)}_\mathrm{BB}\right), {\dots }, R_M\left(\mathbf{w}^{(\kappa)}, \mathbf{F}^{(\kappa)}_\mathrm{RF}, \mathbf{F}^{(\kappa)}_\mathrm{BB}\right)\right) -f\left(R_1\left(\mathbf{w}^{(\kappa)}, \mathbf{F}^{(\kappa)}_\mathrm{RF}, \mathbf{F}^{(\kappa)}_\mathrm{BB}\right), {\dots }, R_M\left(\mathbf{w}^{(\kappa)}, \mathbf{F}^{(\kappa)}_\mathrm{RF}, \mathbf{F}^{(\kappa)}_\mathrm{BB}\right)\right) \nonumber\\ 
&\qquad\qquad\qquad\qquad\qquad\qquad\qquad\qquad\qquad\qquad\qquad\qquad\qquad\qquad\qquad\qquad\times \frac {1}{M}\sum_{m=1}^{M}\frac {R_m\left(\mathbf{w}, \mathbf{F}_\mathrm{RF}, \mathbf{F}_\mathrm{BB}\right)}{R_m\left(\mathbf{w}^{(\kappa)}, \mathbf{F}^{(\kappa)}_\mathrm{RF}, \mathbf{F}^{(\kappa)}_\mathrm{BB}\right)}.
\end{align}
\normalsize
\hrulefill
%\vspace*{4pt}
\end{figure*}
Due to $f\left(R_1\left(\mathbf{w}_1^{(\kappa)}, \mathbf{F}^{(\kappa)}_\mathrm{RF}, \mathbf{F}^{(\kappa)}_\mathrm{BB}\right), {\dots }, R_M\left(\mathbf{w}^{(\kappa)}_M, \mathbf{F}^{(\kappa)}_\mathrm{RF}, \mathbf{F}^{(\kappa)}_\mathrm{BB}\right)\right)>0$, we employ the steepest descent optimization for the convex function $f(R_1, \hdots,R_M)$ to generate the next feasible point $(\mathbf{w}^{(\kappa+1)}, \mathbf{F}^{(\kappa+1)}_\mathrm{RF}, \mathbf{F}^{(\kappa+1)}_\mathrm{BB})$ as follows
\begin{equation}\label{eqn:GM_3}
\begin{aligned}
&\max_{\mathbf{w}, \mathbf{F}_\mathrm{RF}, \mathbf{F}_\mathrm{BB}} \frac {1}{M}\sum_{m=1}^{M}\frac {R_m(\mathbf{w}, \mathbf{F}_\mathrm{RF}, \mathbf{F}_\mathrm{BB})}{R_m\left(\mathbf{w}^{(\kappa)}, \mathbf{F}^{(\kappa)}_\mathrm{RF}, \mathbf{F}^{(\kappa)}_\mathrm{BB}\right)} \\
& \times f\left(R_1(\mathbf{w}^{(\kappa)}, \mathbf{F}^{(\kappa)}_\mathrm{RF}, \mathbf{F}^{(\kappa)}_\mathrm{BB}), {\dots },R_M\left( \mathbf{w}^{(\kappa)}, \mathbf{F}^{(\kappa)}_\mathrm{RF}, \mathbf{F}^{(\kappa)}_\mathrm{BB}\right)\right) \\
& \text {s. t.}\quad \text{(\ref{constr:RBPS_1}), (\ref{constr:RF_1}), (\ref{constr:comb_1}) and (\ref{constr:TP_1})}.
\end{aligned}
\end{equation}
Moreover, the above problem (\ref{eqn:GM_3}) can be equivalently rewritten as
%\vspace{-4mm}
\begin{equation}\label{eqn:GM_4}
\begin{aligned} 
&\hspace{-2mm}\max_{\mathbf{w}, \mathbf{F}_\mathrm{RF}, \mathbf{F}_\mathrm{BB}} \hspace{-2mm}f^{(\kappa)}\left(\mathbf{w}, \mathbf{F}_\mathrm{RF}, \mathbf{F}_\mathrm{BB}\right)\triangleq \sum_{m=1}^{M}\rho^{(\kappa)}_m R_m(\mathbf{w}, \mathbf{F}_\mathrm{RF}, \mathbf{F}_\mathrm{BB})\\ %\label{eqn:eqn_OF_1}\\
& \qquad \text {s. t.}\quad \text{(\ref{constr:RBPS_1}), (\ref{constr:RF_1}), (\ref{constr:comb_1}) and (\ref{constr:TP_1})},
\end{aligned} 
\end{equation}
where $\rho^{(\kappa)}_m$ is given by
\begin{equation}
\begin{aligned}\label{eqn:GM_rho}
\rho^{(\kappa)}_m\triangleq \frac {f\left(R_{1}\left(\mathbf{w}^{(\kappa)}, \mathbf{F}^{(\kappa)}_\mathrm{RF}, \mathbf{F}^{(\kappa)}_\mathrm{BB}\right), \hdots, R_M\left(\mathbf{w}^{(\kappa)}, \mathbf{F}^{(\kappa)}_\mathrm{RF}, \mathbf{F}^{(\kappa)}_\mathrm{BB}\right)\right)}{R_m\left(\mathbf{w}^{(\kappa)}, \mathbf{F}^{(\kappa)}_\mathrm{RF}, \mathbf{F}^{(\kappa)}_\mathrm{BB}\right)}. 
\end{aligned}
\end{equation}
Observe that the problem (\ref{eqn:GM_4}) is a weighted sum-SE maximization, which is similar to (\ref{eqn:system optimization}). Thus, we adopt the sequential approach to optimize the next feasible point $(\mathbf{w}^{(\kappa+1)}, \mathbf{F}^{(\kappa+1)}_\mathrm{RF}, \mathbf{F}^{(\kappa+1)}_\mathrm{BB})$ via solving (\ref{eqn:GM_4}). However, due to the weighing factor $\rho^{\kappa}_m$ in (\ref{eqn:GM_4}), there is modification required in the proposed sequential approach, which is described next.
\begin{enumerate}
    \item To optimize the RF TPC $\mathbf{F}^{(\kappa+1)}_\mathrm{RF}$ and baseband TPC $ \mathbf{F}^{(\kappa+1)}_\mathrm{BB})$, we employ the two-stage hybrid TPC design as discussed in section \ref{hybrid precoder and optimal power allocation}. As a result, the hybrid TPC design problem in the first-stage is modified due to the weighing factor $\rho^{(\kappa)}_m$ as follows
    \begin{equation}\label{eqn:GM_hybrid_TPC_1}
    \begin{aligned}
    \left(\mathbf{F}^{(\kappa+1)}_\mathrm{RF}, \widetilde{\mathbf{F}}_\mathrm{BB}^{1^{(\kappa+1)}}\right) = &\mathop{\arg \min}_{\mathbf{F}_\mathrm{RF}, \widetilde{\mathbf{F}}^{\mathrm{C}, 1}_\mathrm{BB}}\|\mathbf{F}^\mathrm{opt}\mathbf{P}^{(\kappa)}-\mathbf{F}_\mathrm{RF}\widetilde{\mathbf{F}}^{\mathrm{C}, 1}_\mathrm{BB}\|^2_F \\ 
    & \text {s.~ t.} \quad \text{(\ref{constr:RBPS_2}), (\ref{constr:RF_2}) and (\ref{constr:TP_2}),}
    \end{aligned}
    \end{equation}
where $\mathbf{P}^{(\kappa)}$ represents the weighted gain matrix in the $k$th iteration, which is given by $\mathbf{P}^{(\kappa)}= \mathrm{diag}\left(\rho^{(\kappa)}_1, \hdots, \rho^{(\kappa)}_M \right)\in \mathbb{C}^{M \times M}$. Consequently, we adopt the RMCG algorithm to optimize RF TPC $\mathbf{F}_\mathrm{RF}$ except a change in evaluating the Euclidean gradient of $f\left(\mathbf{F}_\mathrm{RF}\right)$ as follows
\begin{equation}\label{eqn:GM_RCG_3}
\begin{aligned}
&\nabla f(\mathbf{F}^{(\kappa+1)}_\mathrm{RF})=\\
&-2\left(\mathbf{F}^\mathrm{opt}\mathbf{P}^{(\kappa)}\left(\widetilde{\mathbf{F}}^{1^{(\kappa)}}_\mathrm{BB}\right)^H + \mathbf{F}^{(\kappa+1)}_\mathrm{RF}\widetilde{\mathbf{F}}^{1^{(\kappa)}}_\mathrm{BB}\left(\widetilde{\mathbf{F}}^{1^{(\kappa)}}_\mathrm{BB}\right)^H\right).
\end{aligned}
\end{equation}
\item For the given RF TPC $\mathbf{F}^{(\kappa+1)}_\mathrm{RF}$, we update $\widetilde{\mathbf{F}}^{1^{(\kappa+1)}}_\mathrm{BB}$ via the least squares as follows
\begin{equation}\label{eqn:GM_F_BB1_tilde}
\begin{aligned}
\widetilde{\mathbf{F}}^{1^{(\kappa+1)}}_\mathrm{BB}=\left(\left(\mathbf{F}_\mathrm{ RF}^{(\kappa+1)}\right)^H\mathbf{F}^{(\kappa+1)}_\mathrm{RF}\right)^{-1}\hspace{-2.5mm}\left(\mathbf{F}_\mathrm{ RF}^{(\kappa+1)}\right)^H\mathbf{F}^\mathrm{opt}\mathbf{P}^{(\kappa)}.
\end{aligned}
\end{equation}
In the second-stage, we evaluate the matrices $\widetilde{\mathbf{F}}^{2^{(\kappa+1)}}_\mathrm{BB}$ and $\mathbf{F}^{2^{(\kappa+1)}}_\mathrm{BB}$ via the ZF and NSP techniques as given by (\ref{eqn:F_BB2_tilde}) and (\ref{eqn:F_BB2_1}), respectively. Then, we update the constrained baseband TPC $\widehat{\mathbf{F}}^{(\kappa+1)}_\mathrm{BB}$ via the RCG method by solving (\ref{eqn:F_BB2_2}).
\item Subsequently, we optimize the RC $\mathbf{w}^{(\kappa+1)}$ by employing the LMBC algorithm and then obtain the rate of each CU $R_m\left(\mathbf{w}^{(\kappa+1)}, \mathbf{F}^{(\kappa+1)}_\mathrm{RF}, \mathbf{F}^{(\kappa+1)}_\mathrm{BB}\right)$. Finally we update the weighing factor $\rho^{(\kappa+1)}_m, \forall m,$ by using (\ref{eqn:GM_rho}).
\end{enumerate}

Since the GM-SE method is also implemented via the sequential approach. Hence, the computational complexity for the GM-SE method is the same as that of Algorithm \ref{alg:algo_3}.

\begin{table}[t]
    \centering
    \caption{Key simulation parameters} \label{tab:simulation parameters}
    %\vspace{-2mm}
    \begin{adjustbox}{width=\linewidth}
\begin{tabular}{l r}\label{Table2}\\
\hline
Parameter & Value \\ 
 \hline
 Carrier frequency & $28$ GHz\\
 Number of propagation paths, $N_m^{\mathrm p}$ & 10\\
 Path-loss parameters & $\varepsilon=61.4, \psi=2, \sigma_{\rm \varpi}=5.8$ dB \cite{HBF_8,ISAC_mmWave_10,ISAC_mmWave_35}\\
 Noise power, $\sigma^2$ & $-91$ dBm\\
 Number of transmit antennas, $N_\mathrm{t}$  & $64$\\
 Number of receive antennas, $N_\mathrm{r}$  & $8$\\
 Number of CUs, $M$ & $3$ \\
 Location of CUs, &$\left[0^\circ, 30^\circ, 60^\circ\right]$ \\
 Number of RTs, $L$ & $2$ \\
 Location of RTs, &$\left[-60^\circ, -20^\circ\right]$ \\
 Angular spread, ${\sigma}_{\theta}$  &$\frac{1}{\sqrt{2}}$\\
 RBPS similarity threshold $\epsilon$ & $\{-30, -10\}$ dB \\
 $\mathrm{SNR}$, $\frac{P_\mathrm{t}}{\sigma^2}$ &  $20$ dB \\
 \hline
\end{tabular}
\end{adjustbox}
%\vspace{-4mm}
\end{table}

%\begin{figure*}[t]
%\centering
%\begin{subfigure}{.65\columnwidth}
%\includegraphics[width=1.1\columnwidth]{R7.eps}%
%\caption{}
%\label{fig:R7}
%\end{subfigure}%\hfill%
%\begin{subfigure}{.65\columnwidth}
%\includegraphics[width=1.1\columnwidth]{R9.eps}%
%\caption{}
%\label{fig:R9}
%\end{subfigure}%
%\begin{subfigure}{.65\columnwidth}
%\includegraphics[width=1.1\columnwidth]{R8.eps}%
%\caption{}
%\label{fig:R8}
%\end{subfigure}%
%\caption{(a) Sum-SE versus number of RAs $N_\mathrm{r}$; (b) Sum-SE versus number of RTs $L$; (c) Sum-SE versus number of CUs $M$.} 
%\vspace{-5mm}
%\end{figure*}
\section{Simulation Results}\label{simulation results}
In this section, we present our simulation results for characterizing the performance of our proposed hybrid TPC/RC aided mmWave MIMO ISAC system operating at the carrier frequency of $28$ GHz. 
The ISAC BS is assumed to have a ULA equipped with $N_\mathrm{t}$ antennas and $M_\mathrm{t}=K$ RF chains. In a similar fashion, each CU has $N_\mathrm{r}$ antennas and a single RF chain located at $\left[0^\circ, 30^\circ, 60^\circ\right]$. The pathloss model $PL(d_m)$ of the mmWave MIMO channel is given by \cite{HBF_8,ISAC_mmWave_10,ISAC_mmWave_35}
\begin{equation}\label{eqn:path loss model}
\begin{aligned}
PL(d_m)\hspace{0.02in}[\rm dB] = \varepsilon + 10\varphi\log_{10}(d_m)+\varpi.
\end{aligned}
\end{equation}
At the carrier frequency of $28$ GHz, we set $\varpi \in {\cal CN}(0,\sigma_{\rm \varpi}^2)$ with $\sigma_{\rm \varpi}=5.8 \hspace{0.02 in}{\rm dB}$, $\varepsilon=61.4$ and $\varphi=2$ for LoS \cite{HBF_8,ISAC_mmWave_10,ISAC_mmWave_35}. 
Additionally, we fix the number of multi-path components to $N_m^{\rm p}=10, \forall m,$ with an angular spread of 10 degrees. Moreover, the AoD and AoA pairs $\theta_{m,l}$ and $\phi_{m,l}, \forall m, l$ are generated from a Laplacian distribution and distributed uniformly within $\left[-90^\circ, 90^\circ\right]$. There are two RTs located at $\overline{\theta }_i \in \left[-60^\circ, -20^\circ\right]$, which have to be identified. Therefore, the desired beampattern is given by
\begin{equation}\label{eqn:desired_beam}
G_{\mathrm{d}}(\theta_{l})= {\begin{cases}1, \theta_{l}\in (\overline{\theta }_i\pm\sigma_{\theta}),i=1,\cdots \!,L,\\ 0,\text{otherwise},
\end{cases}}
\end{equation}
where $\overline{\theta}$ is the mean RT and $\sigma_{\theta}$ denotes a constant angular spread of $\sigma_\theta$, which is assumed to be $\frac{1}{\sqrt{2}}$. It is worth noting here that due to the large number of antenna elements, the transmit beam of the system under consideration is pencil-sharp, and the array response vectors become asymptotically orthogonal even if the angles are very close to each other. Since we assume that the RTs and CUs are sufficiently apart in the angular domain, hence no pairs of transmit beams coincide.
The noise variance $\sigma^2$ at each CU is set to $-91\hspace{0.02 in}{\rm dBm}$. The $\mathrm{SNR}$ is defined as ${\rm SNR}=\frac{P_{\rm t}}{\sigma^2}$, which is set to be the same for every scheme to ensure fairness in the comparisons. 
We compare the performance of our proposed TPC/RC design to that of the following schemes when the similarity thresholds are $\epsilon=-30$ dB and $-10$ dB.
\begin{itemize}
\item FDB-Com only: For this scheme, we assume that the ISAC BS is employed with the fully-digital beamformer (FDB), which serves only the CUs.
\item FRMO-MU: The ISAC BS relies on a hybrid TPC using the fast Riemannian manifold optimization (FRMO) method proposed in \cite{ISAC_mmWave_12} for multiple users, followed by our proposed LMBC algorithm to design the RC at each CU.
\item APSO-MU: For this scheme, a hybrid TPC is employed at the ISAC BS using the adaptive particle swarm optimization (APSO) algorithm proposed in \cite{ISAC_mmWave_12}, followed again by our proposed LMBC algorithm to design the RC at each CU.
\item Two-stage WMMSE: In this method, we harness the two-stage weighted MSE minimization proposed in \cite{ISAC_mmWave_15} for the partially connected mmWave MIMO architecture in order to design the hybrid TPC/RC.
\end{itemize}
%We compare the performance by evaluating the sum-SE of the system versus several important parameters, which are discussed next. 
Unless otherwise stated, we consider an $8 \times 64$ mmWave MIMO system, where the ISAC BS having $N_\mathrm{t} = 64$ antennas and $M_\mathrm{t}=5$ RF chains communicate with CUs that have $N_\mathrm{r} =8$ antennas as well as a single RF chain and detect RTs at a fixed value of SNR $=20$ dB. The key simulation parameters are listed in Table \ref{Table2}. %\textcolor{red}{The simulations are performed on a computer with configuration Intel i7-7700
%CPU  with 16.0 GB RAM operating at 3.60 GHz and }
The simulation curves are averaged over 500 channel realizations to average the outputs of the proposed schemes.
\begin{figure}[t]
    \centering
    \includegraphics[width=0.8\linewidth]{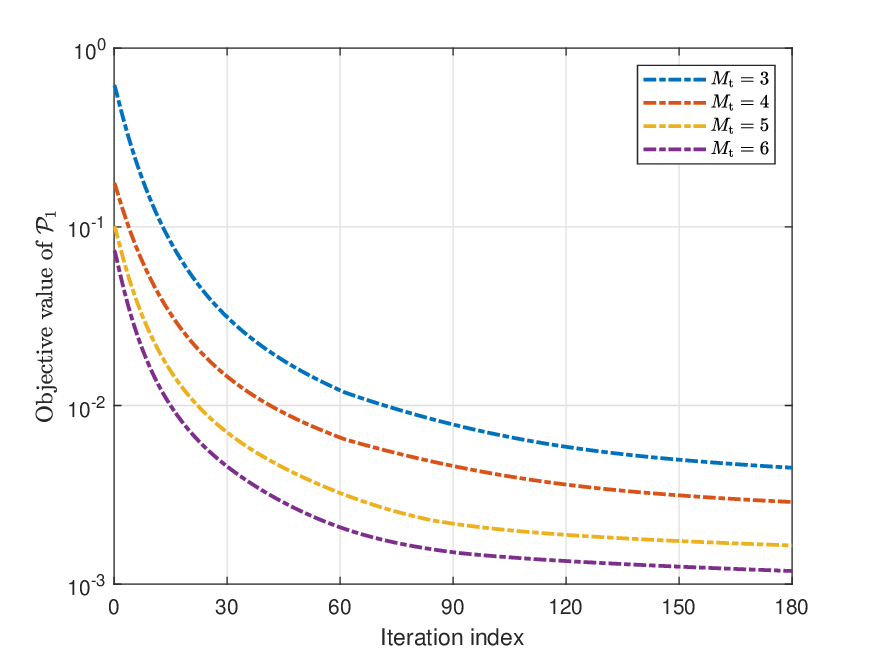}
    \caption{Convergence of the RMCG algorithm for different numbers of RF chains.}
    \label{R12}
\end{figure}
\begin{figure}[t]
    \centering
    \includegraphics[width=0.8\linewidth]{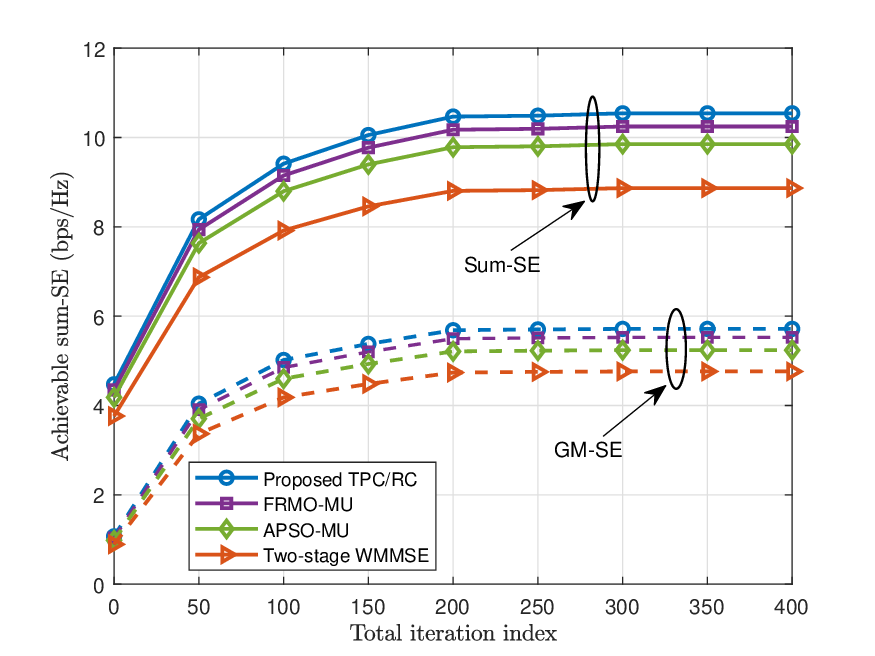}
    \caption{Convergence comparison of the proposed approach with different schemes.}
    \label{R13}
\end{figure}
%\begin{figure}[t]
%\setkeys{Gin}{width=0.8\linewidth}
  %  \begin{subfigure}[t]{0.24\textwidth}
  %  \includegraphics[width=1.1\textwidth]{R12.eps}
 %   \caption{} \label{R12}
%\end{subfigure}
%\begin{subfigure}[t]{0.24\textwidth}
 %   \includegraphics[width=1.1\textwidth]{R13}
 %   \caption{} \label{R13}
%\end{subfigure}
%\caption{(a) Convergence of the RMCG algorithm; (b) Convergence comparison of the proposed approach with different schemes.}
%\vspace{-5mm}
%\end{figure}

%\begin{figure*}[t]%
%\setkeys{Gin}{width=0.8\linewidth}
%\begin{subfigure}{0.25\textwidth}
%\includegraphics{R3.eps}%
%\caption{%Sum-SE versus $\mathrm{SNR}$.
%}
%\label{fig:R3}
%\end{subfigure}%
%\begin{subfigure}{0.25\textwidth}
%\includegraphics{R5.eps}%
%\caption{%Sum-SE versus similarity threshold $\epsilon$.
%}
%\label{fig:R5}
%\end{subfigure}%
%%\label{fig:hasil}
%\begin{subfigure}{0.25\textwidth}
%\includegraphics{R9.eps}%
%\caption{}
%\label{fig:R9}
%\end{subfigure}%
%\begin{subfigure}{0.25\textwidth}
%\includegraphics{R15.eps}%
%\caption{}
%\label{fig:R8}
%\end{subfigure}%
%\caption{Sum-SE versus (a) $\mathrm{SNR}$; (b) similarity threshold $\epsilon$; (c) number of RTs $L$; (d) number of CUs $M$.}
%\vspace{-5mm}
%\end{figure*}
Fig. \ref{R12} shows the convergence behavior of the proposed RMCG algorithm by plotting the objective function value of $\mathcal{P}_2$ versus the iteration number for different numbers of RF chains. It can be seen from the figure that the objective value of $\mathcal{P}_2$ decreases with the iteration index for all $M_\mathrm{t}$ and converges after $170$ iterations. Observe that the decrement rate of the objective value of $\mathcal{P}_2$ increases with increases in $M_\mathrm{t}$. This is because large $M_\mathrm{t}$ results in the better approximation of hybrid TPC in the first-stage with the optimal TPC.
Furthermore, Fig. \ref{R13} compares the convergence behavior of the overall proposed TPC/RC scheme for both sum-SE and GM-SE to the state-of-art schemes by plotting the achievable sum-SE versus the maximum number of iterations along with $M_\mathrm{t}=5$. As seen from the figure, the achievable sum-SE of the proposed TPC/RC design converges after 250 iterations.

\begin{figure}[t]
    \centering
    \includegraphics[width=0.8\linewidth]{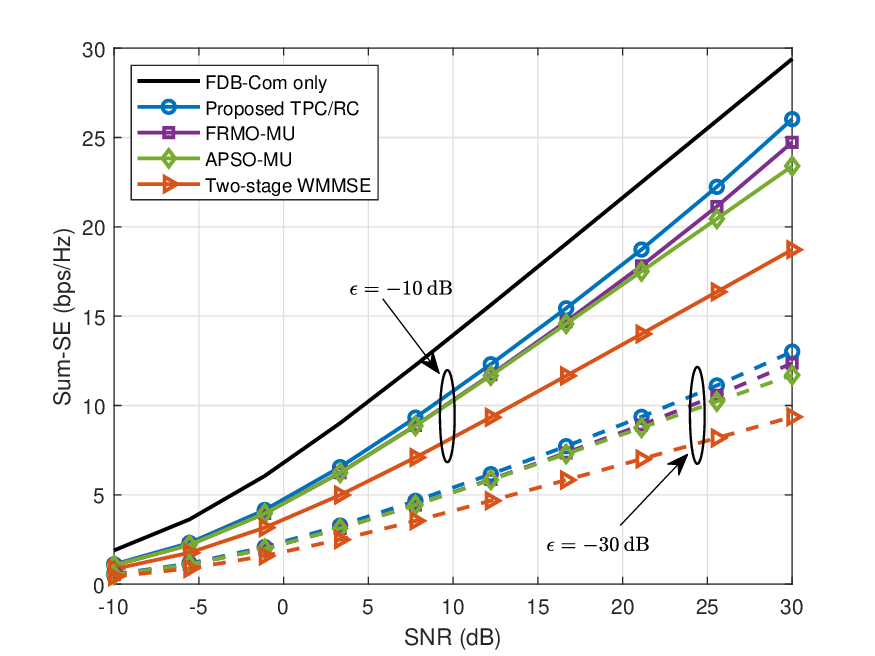}
    \caption{Sum-SE versus $\mathrm{SNR}$}
    \label{fig:R3}
\end{figure}
To show the trade-off between the radar and communication performance achieved by the hybrid TPC/RC, we compare the sum-SE of the system versus the SNR, when the similarity threshold is $\epsilon=\{-30, -10\}$ dB in Fig. \ref{fig:R3}. As seen from the figure, the sum-SE of the proposed hybrid TPC/RC design increases with the SNR due to the improvement of the resultant SINR at the CUs with a higher transmit power. Furthermore, the sum-SE of the proposed hybrid TPC/RC design
approaches that of the optimal-FDB design for $\epsilon=-10$ dB and performs poorly at $\epsilon=-30$ dB. This is because as the MSE requirement of the RBPS decreases, which increases the emphasis on the RTs, it results in a sum-SE reduction for the system. Hence, the similarity threshold $\epsilon$ controls the trade-off between the communication and radar performance. Furthermore, the performance of the proposed hybrid TPC/RC design is better than that of the state-of-the-art techniques, which shows the benefits of our proposed RMCG and LMBC algorithms. This is due to the fact that state-of-the-art techniques rely on the hybrid TPC by minimizing the weighted sum of the communication and radar beamforming errors. By contrast, our proposed RMCG algorithm optimizes the hybrid TPC, which maximizes the sum-SE under a specific MSE constraint. 

%\begin{figure}
%\centering
%\includegraphics[width = 8.5 cm]{R5.eps}
%\caption{Sum-SE versus similarity threshold $\epsilon$ of the proposed hybrid TPC/RC design for an $8\times 64$ mmWave MIMO ISAC system.}
%\label{fig:R5}
%\vspace{-4mm}
%\end{figure}
\begin{figure}[t]
    \centering
    \includegraphics[width=0.8\linewidth]{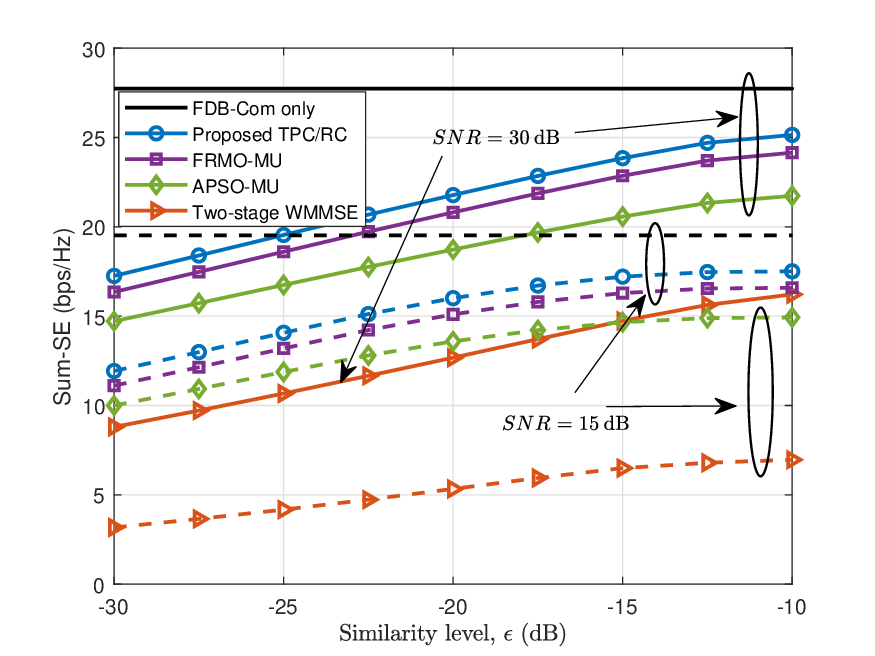}
    \caption{Sum-SE versus similarity threshold $\epsilon$.}
    \label{fig:R5}
\end{figure}
To further evaluate the performance of the proposed hybrid TPC/RC design, we compare the sum-SE of the system with respect to the RBPS threshold $\epsilon$ for SNR $\in \{15, 30\}$ dB in Fig. \ref{fig:R5}. As seen from the figure, the sum-SE of the system improves upon increasing the RBPS threshold $\epsilon$. This is due to the fact that a large value of $\epsilon$ provides an opportunity for the ISAC BS to focus its beams more towards the CUs due to the improved capability of the RTs to tolerate the MSE of the RBPS. Therefore, one can optimize the transmit power at the ISAC BS to achieve a particular sum-SE of the system for a given MSE requirement of the RTs. Moreover, the proposed design performs better than other schemes at low as well as higher values of $\epsilon$ and at both SNR= $15$ dB and $30$ dB, which shows the efficiency of our RMCG and LMBC algorithms in maximizing the sum-SE of the system.    
\begin{figure}[t]
    \centering
    \includegraphics[width=0.8\linewidth]{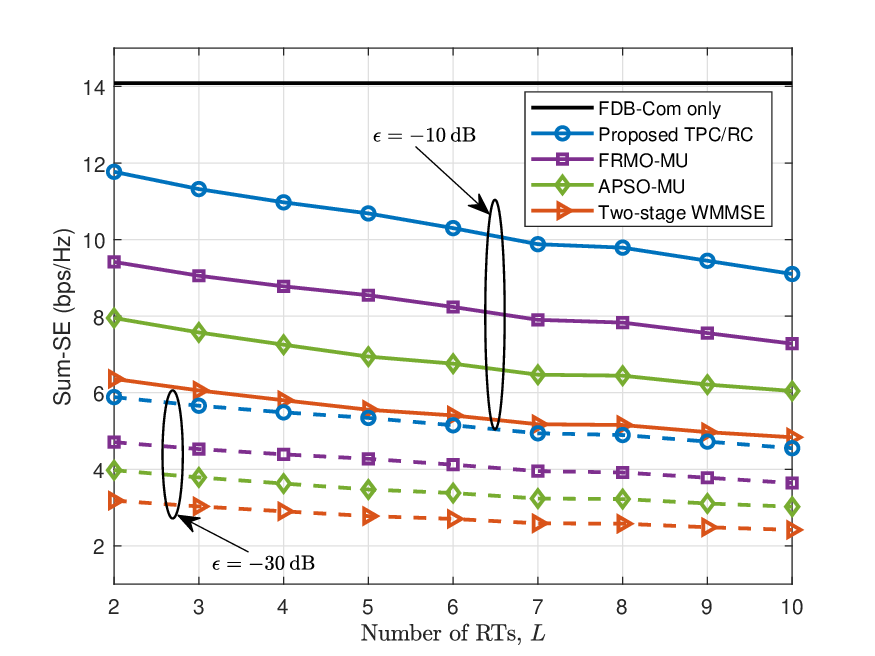}
    \caption{Sum-SE versus number of RTs $L$.}
    \label{fig:R9}
\end{figure}
To further explore the associated performance trade-off between radar and communication performance, in Fig. \ref{fig:R9}, we investigate the sum-SE of the system versus the number of RTs $L$ for a fixed value of $\mathrm{SNR=20}$ dB and $M=3$. In this case, we set the locations of the RTs in the angular bin of $\left[-90^\circ, 0^\circ\right]$, where each RT is separated with a gap of $10^\circ$.
Observe from the figure that the performance of the system degrades upon increasing the number of targets. This is due to the fact that as the number of RTs increases, the ISAC BS has to form distinct transmit beams towards each of them. Hence, the limited transmit power is distributed among the incoming RTs, which results in an SINR reduction at the CUs. Therefore, the sum-SE of the system decreases upon increasing the number of the RTs. Moreover, the optimal-FDB scheme is unaffected by the RTs, as it optimizes the fully digital TPC while considering the CUs in the environment. Moreover, the system performance is improved upon increasing the similarity threshold from $\epsilon=-30$ dB to $\epsilon=-10$ dB, demonstrating that relaxing the MSE threshold leads to accommodating more RTs in the system.
%\begin{figure}
%\centering
%\includegraphics[width = 8.5 cm]{R8.eps}
%\caption{Sum-SE versus number of CUs $M$ of the proposed hybrid TPC/RC design for an $8\times 64$ mmWave MIMO ISAC system with $\mathrm{SNR}=20$ dB.}
%\label{fig:R8}
%\vspace{-4mm}
%\end{figure}

\begin{figure}[t]
    \centering
    \includegraphics[width=0.8\linewidth]{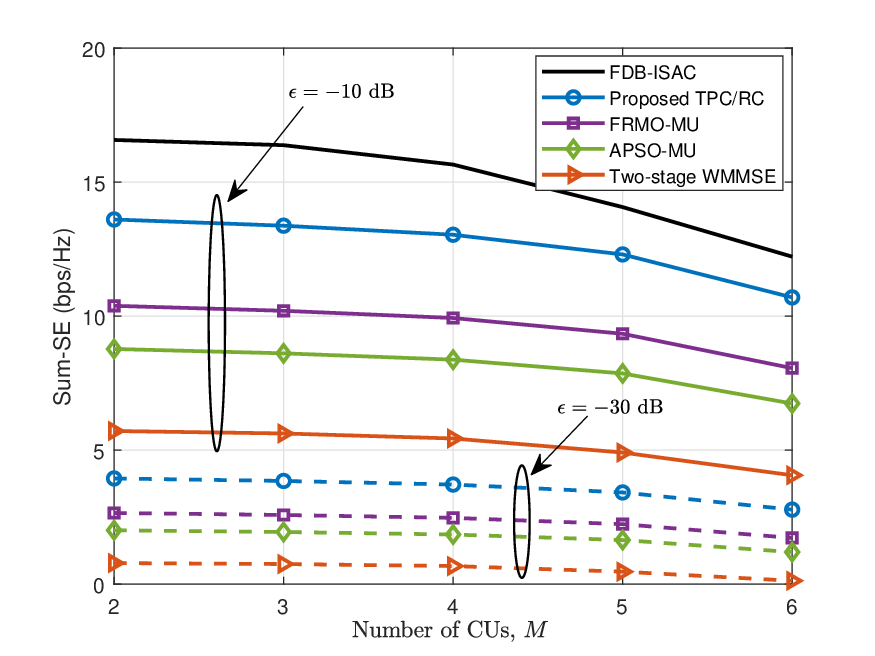}
    \caption{Sum-SE versus number of CUs $M$.}
    \label{fig:R8}
\end{figure}
In Fig. \ref{fig:R8}, we investigate the sum-SE versus the number of CUs $M$ for SNR$=20$ dB and the number of RTs $L=2$. As seen from the figure, the sum-SE of the system decreases as $M$ increases due to the increment in the MUI and reduction in the power per CU.  To compensate for these losses, it is advisable to increase the number of receive antennas as $M$ increases, since the receive antennas have a higher impact on the sum-SE in comparison to transmit antennas due to the associated MSE constraint.
For a fair comparison, we use the FDB-ISAC benchmark scheme, where the ISAC BS employed the FDB architecture and served the CUs as well as detected the RTs with the available power.
As seen from the figure, the proposed TPC/RC design approaches to the FDB-ISAC scheme for $\epsilon=-10$ dB. This reveals that the proposed TPC/RC design using fewer RF chains performs almost the same as the optimal FDB-ISAC scheme.
Moreover, the proposed TPC/RC scheme outperforms the benchmark schemes as $M$ increases, which shows the efficiency of our two-stage TPC design.

%\begin{figure}
  %      \begin{subfigure}[b]{0.28\textwidth}
  %      \hspace{-6mm}
  %              \includegraphics[width=0.8\linewidth]{R1.eps}
  %              \caption{}
  %              \label{fig:R1}
  %      \end{subfigure}%
  %      \begin{subfigure}[b]{0.28\textwidth}
 %       \hspace{-10mm}
 %               \includegraphics[width=0.8\linewidth]{R2.eps}
 %               \caption{}
%                \label{fig:R2}
%        \end{subfigure}%
%        \caption{Transmit beampattern of a) communication-only and radar-only; b) proposed hybrid TPC/RC design.}
 %       \vspace{-5mm}
%\end{figure}

%\begin{figure}
%\centering
%\includegraphics[width = 7 cm]{R1.eps}
%\caption{Transmit beampattern of the communication and radar for an $8\times 64$ mmWave MIMO ISAC system with $SNR=20$ dB.}
%\label{fig:R1}
%\vspace{-4mm}
%\end{figure}

%\begin{figure}
%\centering
%\includegraphics[width = 7 cm]{R2.eps}
%\caption{Transmit beampattern of the proposed hybrid TPC/RC design for an $8\times 64$ mmWave MIMO ISAC system with $SNR=20$ dB.}
%\label{fig:R2}
%\vspace{-4mm}
%\end{figure}
\begin{figure}[t]
    \centering
    \includegraphics[width=0.8\linewidth]{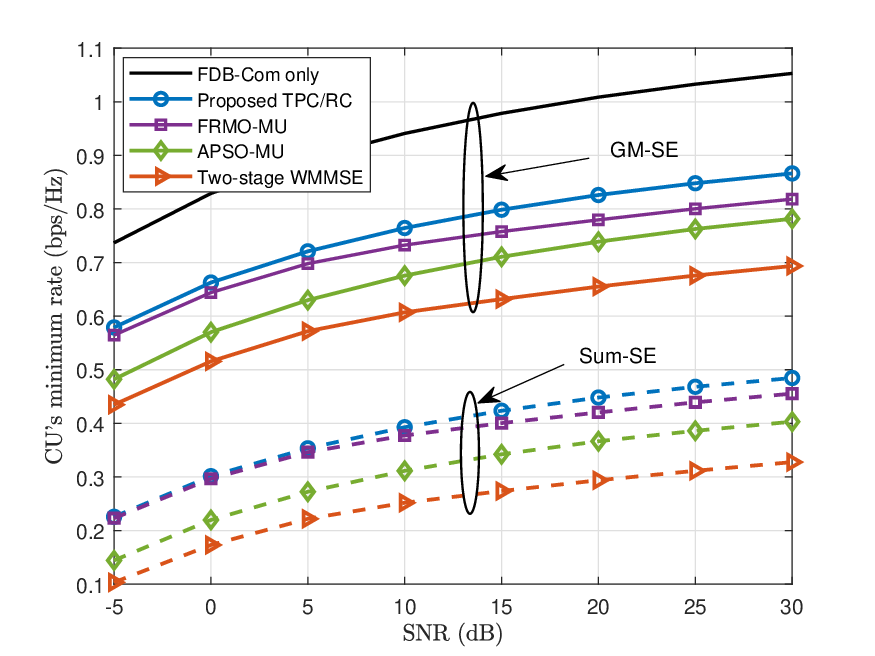}
    \caption{CU's minimum rate versus $\mathrm{SNR}$.}
    \label{fig:R11}
\end{figure}
\begin{figure}[t]
    \centering
    \includegraphics[width=0.8\linewidth]{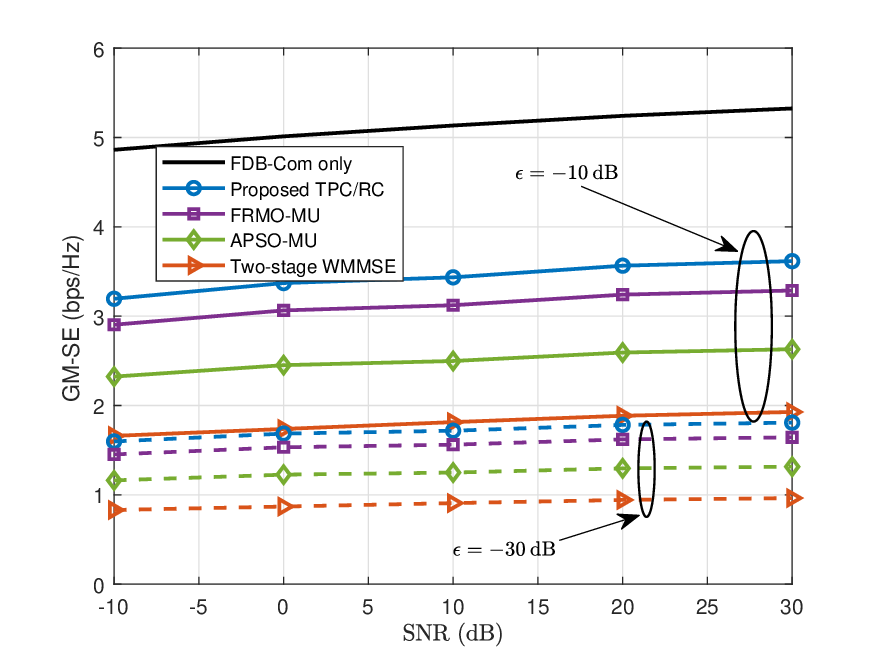}
    \caption{GM-SE versus $\mathrm{SNR}$.}
    \label{fig:R10}
\end{figure}
\begin{figure}[t]
    \centering
    \includegraphics[width=0.8\linewidth]{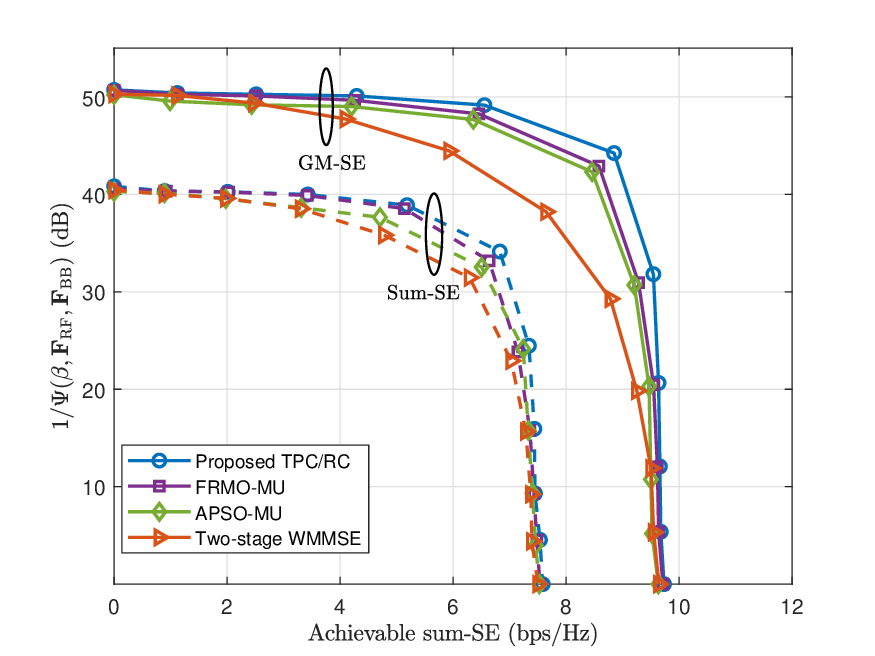}
    \caption{Trade-off between sensing and communication performance.}
    \label{fig:R14}
\end{figure}
Since the proposed TPC/RC design based on the sum-SE maximization results in the CU's rate unfairness, we investigate the user fairness by plotting the GM-SE vs SNR in Fig. \ref{fig:R11} and Fig. \ref{fig:R10}. In Fig. \ref{fig:R11}, we plot the CU's minimum rate versus SNR at $\epsilon=-30$ dB for both the sum-SE and GM-SE metrics. As seen from the figure, GM-SE achieves a better minimum rate than sum-SE, which shows that the GM-SE metric is more favorable than the sum-SE for maintaining rate fairness among the CUs in the system. Furthermore, Fig. 4b reveals the impact of $\epsilon$ on the performance of the GM-SE. As expected, the GM-SE of the system improves upon increasing $\epsilon$ from $-30$ dB to $-10$ dB, as seen in the description of Fig. \ref{fig:R3}.

To investigate the trade-off between sensing and communication performance, Fig. \ref{fig:R14} plots the RBPS of the RTs versus the achievable sum-SE of the system for both the sum-SE and GM-SE methods. As seen from the figure, the sensing performance degrades upon increasing the achievable sum-SE, and hence, there is a trade-off between the sensing and communication performance. Consequently, one can corroborate the beampattern MSE trends also in terms of the detection probability of the RTs. For a given sum-SE, the detection probability increases upon reducing the beampattern MSE $\Psi(\beta,\mathbf{F}_\mathrm{RF}, \mathbf{F}_\mathrm{BB})$ due to the increased gain towards the RTs. Also, observe that the RBPS of our proposed TPC/RC approach is better than the benchmark schemes, which shows the efficacy of the proposed design for the detection of multiple RTs. 
Furthermore, the trade-off boundary of the GM-SE method is better than that of the sum-SE method because of the availability of more power for the GM-SE method.

%For characterizing the tradeoff between the radar
%and communication performance achieved by the proposed TPC/RC design, we plot the resultant transmit beampattern in Fig. \ref{fig:R1} and \ref{fig:R2}. In Fig. \ref{fig:R1}, we plot separate beampatterns for the communication and radar systems. As seen from this figure, the ISAC BS aligns the main lobes toward the CUs in the communication system and toward the RTs for the radar operation. 

\begin{figure}[t]
    \centering
    \includegraphics[width=0.8\linewidth]{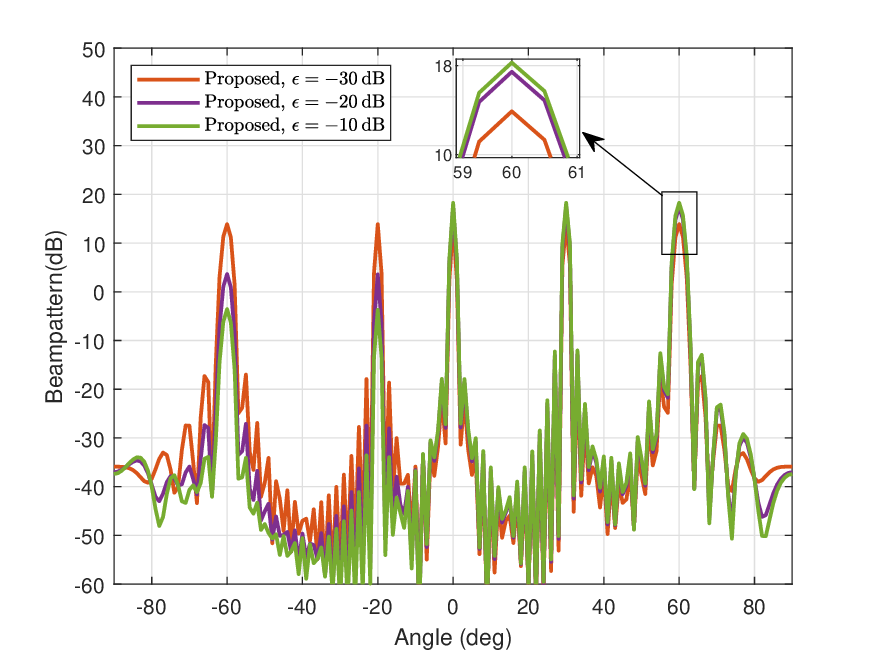}
    \caption{Transmit beampattern.}
    \label{fig:R2}
\end{figure}
\begin{figure}[t]
    \centering
    \includegraphics[width=0.8\linewidth]{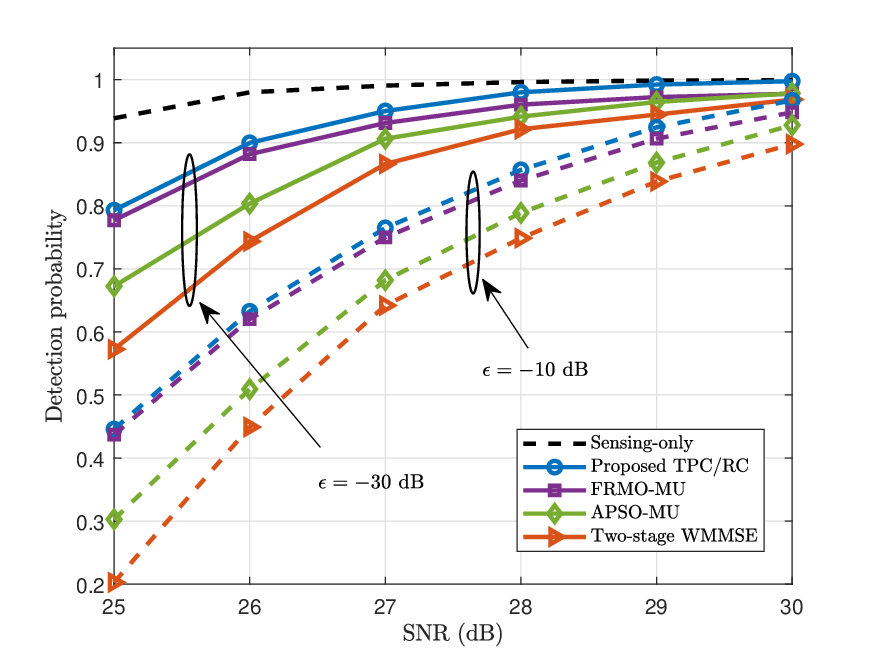}
    \caption{Detection probability versus SNR under false alarm probability $P_\mathrm{FA} = 10^{-4}$ for sum-SE = 15 bps/Hz.}
    \label{fig:R17}
\end{figure}
\begin{figure}[t]
    \centering
    \includegraphics[width=0.8\linewidth]{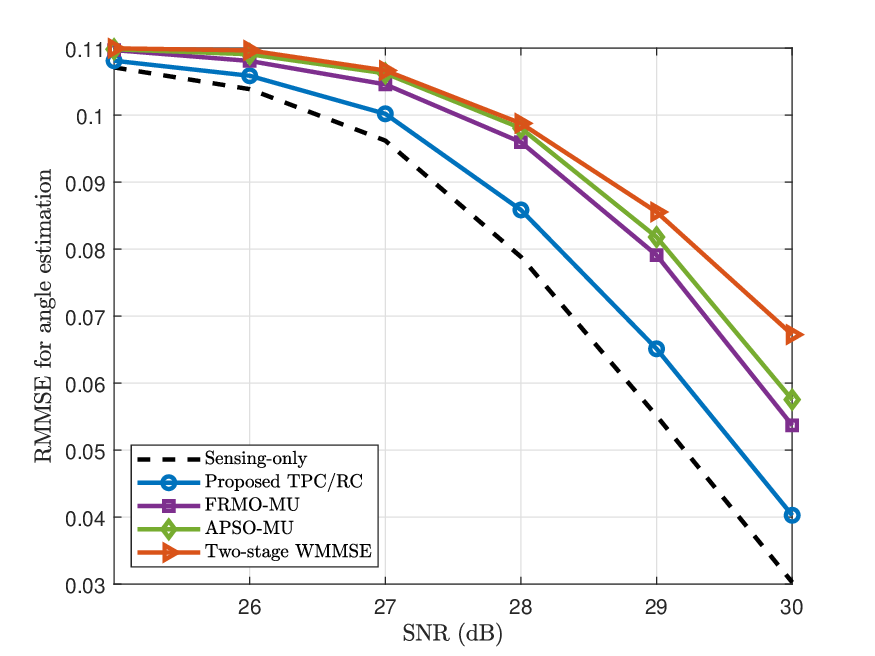}
    \caption{RMSE for angle estimation versus SNR for sum-SE = 15 bps/Hz.}
    \label{fig:R16}
\end{figure}

%\begin{figure*}[t]
%\centering
%\begin{subfigure}{.65\columnwidth}
%\includegraphics[width=1\columnwidth]{R2.eps}%
%\caption{}
%\label{fig:R2}
%\end{subfigure}%\hfill%
%\begin{subfigure}{.65\columnwidth}
%\includegraphics[width=1\columnwidth]{R17.eps}%
%\caption{}
%\label{fig:R17}
%\end{subfigure}%
%\begin{subfigure}{.65\columnwidth}
%\includegraphics[width=1\columnwidth]{R16.eps}%
%\caption{}
%\label{fig:R16}
%\end{subfigure}%
%\caption{(a) Transmit beampattern; (b) Detection probability versus SNR under false alarm
%probability $P_\mathrm{FA} = 10^{-4}$ for sum-SE = 15 bps/Hz; (c) RMSE for angle estimation versus SNR for sum-SE = 15 bps/Hz.}
%\end{figure*}

Fig. \ref{fig:R2} shows the transmit beampattern of the proposed hybrid TPC/RC design. Observe that the ISAC BS is able to focus the lobes toward the CUs as well as the RTs. 
This observation highlights the ability of the proposed TPC/RC method to resolve multiple RTs by suppressing the interference due to multiple RTs through the RBPS constraint.
Furthermore, the gain of the beampattern towards the RTs increases with the reduction in MSE threshold from $\epsilon=-10$ dB to $\epsilon=-30$ dB. This is due to the fact that smaller $\epsilon$ value results in a higher similarity between the ideal and the designed beampatterns, and hence, more power is radiated toward the RTs. %Moreover, note that in the benchmark schemes, one has to optimize the weighting factor between the communication and radar beamforming errors to optimize the ISAC design. By contrast, our proposed scheme does not require such an optimization, which results in robust ISAC systems.

To investigate the sensing performance, we evaluate the detection probability of the RTs using the Generalized Likelihood Ratio Test (GLRT) under a fixed sum-SE requirement for the CUs, as discussed in \cite{new_100}. Fig. \ref{fig:R17} illustrates the detection probability versus SNR for a false alarm probability of $P_\mathrm{FA} = 10^{-4}$ and a fixed sum-SE of $15$ bps/Hz, comparing the proposed techniques with the sensing-only scheme. In the sensing-only scheme, all resources at the ISAC BS, including the power and RF chains, are fully allocated to sensing, serving as an upper bound for RT detection.
As observed, the detection probability improves upon increasing the SNR, and the proposed design closely approaches the optimal sensing-only scheme at higher SNRs due to the improved transmit beampattern gain towards RTs for the given sum-SE. Moreover, as expected, reducing the beampattern similarity threshold from $\epsilon = -10$ dB to $\epsilon = -30$ dB further improves the detection probability. This is because a lower threshold minimizes the MSE between the optimal and designed transmit beampatterns, directing more power towards the RTs. Importantly, our proposed design outperforms the benchmark schemes, demonstrating its effectiveness in enhancing the RT sensing performance.

To further explore the estimation performance, we evaluate the root-mean-square-error (RMSE) of the estimated angle of the RTs. Specifically, the RMSE is formulated as
\begin{equation} 
\mathrm{RMSE} = \sqrt{ {\mathbb{E}} \left\lbrace \frac{1}{L} \sum _{l=1}^{L} (\overline{ \theta }_l - \hat{\theta }_l) ^ 2 \right\rbrace }, 
\end{equation}
where $\overline{\theta }_l$ and $\hat{\theta }_l$ are the true and estimated angles of the $l$th RT, for $l=1, \hdots, L$. Furthermore, we employed the MUSIC algorithm \cite{ISAC_mmWave_1_1} for the estimation of the angles of the RTs. 
Fig. \ref{fig:R16} depicts the RMSE in degree versus SNR for a fixed sum-SE requirement of $15$ bps/Hz. As expected the RMSE decreases upon increasing the SNR due to the resultant higher power radiated towards the RTs. Notably, the proposed design achieves a performance close to the optimal sensing-only scheme, demonstrating its effectiveness in target estimation while simultaneously serving the CUs.

\section{Conclusion}\label{conclusion}
A hybrid TPC and RC pair was designed for maximizing the sum-SE and GM-SE of mmWave MIMO ISAC systems, while meeting a practical RBPS constraint pertaining to the MSE of the radar beampattern gain. A two-stage hybrid TPC design was proposed in which the RF TPC was obtained using the RMCG algorithm, while the baseband TPC was determined using the least squares and ZF methods. Moreover, in order to circumvent the feedback of the TPCs between the ISAC BS and CUs to design the RCs, a low-complexity LMBC algorithm was also proposed. 
Furthermore, simulation results were presented, which show the benefits of the proposed TPC/RC-aided mmWave MIMO ISAC design. Future research ideas include the development of a hybrid TPC/RC design for frequency-selective mmWave MIMO ISAC systems, while satisfying the specific MSE constraint. 
\bibliographystyle{IEEEtran}
\bibliography{biblio.bib}
\end{document}